\RequirePackage{fix-cm}
\documentclass[twocolumn]{svjour3}          %
\smartqed  %
\usepackage{graphicx}

\usepackage{graphicx}%
\usepackage{multirow}%
\usepackage{amsmath, amssymb,amsfonts}%
\usepackage{mathrsfs}%
\usepackage[title]{appendix}%
\usepackage{xcolor}%
\usepackage{textcomp}%
\usepackage{manyfoot}%
\usepackage{booktabs}%
\usepackage{algorithm}%
\usepackage{algorithmicx}%
\usepackage{algpseudocode}%
\usepackage{listings}%
\usepackage{tikz}%
\usetikzlibrary{spy}
\usepackage{makecell} %
\usepackage{enumitem}
\usepackage[numbers,square]{natbib} %
\usepackage[hidelinks]{hyperref}
\usepackage{orcidlink}
\usepackage{overpic}

\usepackage{mathtools}
\usepackage{xspace}
\newcommand{\FLIP}{\protect\reflectbox{F}LIP\xspace}


\DeclareMathOperator*{\argmin}{arg\,min}

\newcommand{\lod}[0]{LoD}

\newcommand{\new}[1]{#1}

\journalname{The Visual Computer}
\begin{document}

\title{Fast Voxelization and Level of Detail for Microgeometry Rendering %
}

\author{Javier Fabre~\orcidlink{0000-0002-0846-8121} \and
        Carlos Castillo \and
        Carlos Rodriguez-Pardo~\orcidlink{0000-0001-6121-7738}  \and
        Jorge Lopez-Moreno~\orcidlink{0000-0001-6278-6940}
}

\institute{
  Javier Fabre \at
    Universidad Rey Juan Carlos \\
    \email{fjavifabre@gmail.com}          
  \and
  Carlos Castillo \at
  Universidad Rey Juan Carlos \\
    \email{carlos.castillo@urjc.es} 
  \and
  Carlos Rodriguez-Pardo \at
    Politecnico di Milano \\
    Euro-Mediterranean Center on Climate Change (CMCC) \\
    RFF-CMCC European Institute on Economics and the Environment (EIEE) \\
    \email{carlos.rodriguezpardo.jimenez@gmail.com}
  \and
  Jorge Lopez-Moreno \at
    Universidad Rey Juan Carlos \\
    \email{jorge.lopez@urjc.es}
}

\date{Received: date / Accepted: date}

\maketitle

\begin{abstract}
    Many materials show anisotropic light scattering patterns due to the shape and local alignment of their underlying micro structures: surfaces with small elements such as fibers, or the ridges of a brushed metal, are very sparse and require a high spatial resolution to be properly represented as a volume.     
    The acquisition of voxel data from such objects is a time and memory-intensive task, and most rendering approaches require an additional Level-of-Detail (LoD) data structure to aggregate the visual appearance, as observed from multiple distances, in order to reduce the number of samples computed per pixel (E.g.: MIP mapping).     
    In this work we introduce first, an efficient parallel voxelization method designed to facilitate fast data aggregation at multiple resolution levels, and second, a novel representation based on hierarchical SGGX clustering that provides better accuracy than baseline methods. 
    We validate our approach with a CUDA-based implementation of the voxelizer, tested both on triangle meshes and volumetric fabrics modeled with explicit fibers. Finally, we show the results generated with a path tracer based on the proposed LoD rendering model.
\end{abstract}

\sloppy

\section{Introduction}\label{sec:introduction}

Voxel-based data representations have become fundamental tools across numerous domains in computer graphics, as well as medical and scientific data. Their structured nature provides an efficient discretization of 3D-spaces which enables complex algorithms to process geometric data systematically. Voxels have been used in rendering for fast ray-tracing~\cite{Amanatides:1987:Fast}, volumetric path-tracing~\cite{Lafortune:1996:Rendering}, shadow computation and visibility~\cite{Aleksandrov:2019:Voxel} among other applications.

Recent advances in neural scene representations, particularly Neural Radiance Fields~\cite{Mildenhall:2020:NeRF, Kaizhang:2020:Nerf++} (\emph{NeRF}), have renewed the interest in efficient volume representation and rendering. While learning-based approaches may offer impressive results, they often come with significant computational complexity and a lack of desirable properties like editability. More recent works, like \emph{Gaussian Splatting} (GS)~\cite{Kerbl:2023:3DGaussians,Wu:2024:4DGaussians} and its ray-tracing counterparts~\cite{Moenne-Loccoz:2024:Gaussian-RT,Condor:2025:DontSplat}, offer different trade-offs between quality, speed and memory use. Building upon these works, approaches like~\emph{Radiant Foam}~\cite{Govindarajan:2025:Radfoam}, or sparse voxels~\cite{Sun:2024:SparseVoxel} have shown comparable or superior performance. The combination of simpler analytic representations with sparse voxel flexibility, as in SplatVoxel~\cite{Wang:2025:Splatvoxel}, is promising, especially for temporal coherence. While RAW data has its uses, complex representations via distributions enable higher information storage per voxel and easier level-of-detail (\lod), also performing well in optimization, as demonstrated by GS.

Despite these advances, a significant challenge remains in efficiently representing and rendering materials with anisotropic microstructures—such as fabrics, hair, and brushed metals. These materials exhibit high sparsity (occupying a small fraction of the volume), require high spatial resolution to capture fine details, and exhibit complex directional properties that define their appearance.

To utilize voxels in rendering, vector-based primitives such as points, lines, splines, triangles, surfaces, and solids must undergo a voxelization process. This transformation can become computationally prohibitive, especially for complex geometry. Consequently, previous research has developed approaches that optimize voxelization for specific primitive types~\cite{Aleksandrov:2021:Voxelisation} or scenarios, such as high-resolution voxel grids~\cite{Schwarz:2010:Fast}, complex models~\cite{Dong:2004:Real-time}, or image-based data~\cite{Loop:2013:Real-time}. While voxelized volumes for rendering are inherently sparse~\cite{Museth:2013:VDB}, microgeometry exhibits particular sparsity. Materials with fine-scale structures—such as fabrics, hair, and brushed metals—present two challenges: they occupy a minimal fraction of the volume while simultaneously requiring high resolution to capture their detail. This combination creates significant computational and memory complexity that existing approaches struggle to address efficiently.

To our knowledge, no prior method specifically addresses the GPU-optimized voxelization of complex data in the highly sparse scenarios characteristic of microgeometry, despite the prevalence of such materials in real-world applications. To fill this gap, we present the following contributions:

\begin{itemize}
    \item \textbf{A novel voxelization method} optimized for sparse microgeometry. Our approach handles various input primitives, including splines and sources with textured orientation data, while maintaining memory efficiency at high resolutions.
   \item \textbf{A GPU-CPU implementation} that maximizes voxelization speed and memory efficiency through strategic workload distribution, enabling processing of highly detailed microgeometry that would exceed memory limits with traditional approaches.
   \item \textbf{SGGX-H}: A novel hierarchical volumetric data model for \lod~rendering based on Symmetrical GGX (SGGX)~\cite{Heitz:2015:SGGX}. Unlike previous approaches that lose critical directional information through simple averaging, our method clusters and preserves directional distributions across resolution levels, maintaining the characteristic anisotropic appearance of complex materials.
\end{itemize}

We validate our approach through a CUDA-based implementation tested on both triangle meshes and volumetric fabrics with explicit fibers, demonstrating significant improvements compared to baseline approaches. Upon publication, we will release the source code and render scenes.

\section{Previous Work}

In the following, we discuss current voxelization methods, the problem of representing and voxelizing micro-geometry, and the state of the art in level-of-detail filtering. 

\subsection{Voxelization methods}

\label{ssec:voxelization-methods}

Before volumetric data can be used, it must be stored in a format allowing both efficient storage and fast access. Data compression is essential for high-resolution volumes, making sparse representations the standard approach. OpenVDB~\cite{Museth:2013:vdbCourse, OpenvdbSoftware} and its GPU counterpart NanoVDB~\cite{Museth:2021:nanovbd,NanoVDBSoftware} are the industry standard for voxel data management, offering excellent performance for both reading and writing operations. However, neither framework focuses on minimizing voxelization time or optimizing for specific primitives. Instead, they provide raster-based voxelization methods for meshes that prioritize storage density over processing speed, primarily using CPU parallelization rather than fully exploiting GPU capabilities for the voxelization process itself.

Instead, researchers have developed specialized approaches for different 3D primitive types, including lines~\cite{Cohen:1997:Line-voxelization}, triangles~\cite{Ix:2000:Incremental}, polygons~\cite{Kaufman:1987:3DScan}, and solids~\cite{Eisemann:2006:Fast}. For more computationally intensive scenarios, Crassin et al.~\cite{Crassin:2011:GigaVoxels,Crassin:2011:Interactive} pioneered GPU-based algorithms for volumetric data generation. For a comprehensive review of voxelization methods and recent advances, we refer readers to the survey by Aleksandrov et al.~\cite{Aleksandrov:2021:Voxelisation}.
Beyond simple voxelization, accurately storing complex properties such as optical density, orientation distributions, or statistical data presents additional challenges. While OpenVDB and NanoVDB support custom data types, operations like down-sampling and interpolation become problematic for these complex properties. Most existing voxelization techniques focus primarily on density values, requiring significant modifications to effectively capture and store directional or statistical information that is crucial for accurately representing anisotropic materials.

\subsection{Voxelizing and rendering micro structures}

Rendering materials with micro-geometric details using explicit geometry or high-resolution orientation and displacement maps presents significant computational and memory challenges. These representations not only demand substantial resources but also frequently suffer from aliasing artifacts, making the efficient representation and rendering of micro-geometry an ongoing challenge in computer graphics.

Volumetric approaches offer a promising alternative, with various specialized scattering models developed for different material types, including microflakes and scattering models for fabrics. Khungurn et al.~\cite{Khungurn:2016:Matching} demonstrated that accurately representing fiber-like materials requires dedicated scattering models tailored to their unique properties. We refer the reader to the survey by Castillo and colleagues~\cite{Castillo:2019:Recent} for further information about the different techniques used for cloth rendering.

Lopez-Moreno et al.~\cite{Lopez-Moreno:2017:Sparse} adapted the \textit{Lumislice}~\cite{Ying:2001:Lumislice} algorithm for volumetric cloth representation in GPU\@. Their approach deals with some of the problems of fiber-like materials; however, it is not suitable for \lod~generation, as samples are generated directly at the finer level, with simple averaging of multiple fragments per voxel. This operation fails to preserve critical orientation data, making accurate scattering simulation impossible at coarser detail levels.

Our voxelization method addresses this limitation, preserving orientation distributions across multiple detail levels. By maintaining accurate directional information even at lower resolution levels, fibers and textured surfaces can be rendered with less samples, using various anisotropic scattering models while preserving their characteristic appearance across different viewing distances.

\subsection{Filtering and Level-of-Detail}

MIP-mapping~\cite{Williams:1983:Pyramidal} has been widely use since it was presented for textures \lod. Usually, the MIP-map data is generated by linearly interpolating pyramid levels.
This approach works for some parameters like \textit{albedo} or \textit{roughness}, but fails for \textit{alpha} or orientation data (i.e. \textit{normal} or \textit{tangent maps}).

Modifying additional maps can solve this issue.
LEAN-Mapping~\cite{Olano:2010:LEAN} has been  used for real-time scenarios to overcome the problems that arise from normal-map interpolation.
Gauthier and colleagues~\cite{Gauthier:2022:MIPNet} addressed the issue of normal MipMapping with a neural approach, also showing that straight-forward down-sampling of normal-maps requires changes in \textit{roughness} for each \lod.
These techniques do not generate correct input data, instead relying in the material to visually fix the wrong appearance that the data generates in the final render. It would be beneficial to obtain a representation of the material properties that is agnostic to the material model, so different models could be used to render such data without any modifications.

Mesh decimation has been used to generate different \lod~meshes that can be used to save space when rendering complex. Commonly, these new meshes for \lod~are generated using either \textit{edge contraction}~\cite{Garland:1997:Surface}, \textit{vertex decimation}~\cite{Schroeder:1997:Topology} or vertex clustering~\cite{Low:1997:Model} techniques driven by some metric to measure quality of the resulting mesh for a given \lod. However, these metrics often failed to accurately represent all factors that influence \lod~appearances. Recently,  there has been major advances in inverse rendering techniques (please refer to Laine et al.~\cite{Laine:2020:Modular} for an overview of current approaches) that help generate new metrics that can account for final appearance to drive mesh optimization, such as the proposed by Hasselgren and colleagues~\cite{Hasselgren:2021:Appearance}. 

Previous work in~\lod~generation has followed two main approaches: either optimizing each level to minimize memory consumption~\cite{Haydel:2023:Locally}, thereby reducing overall asset size, or optimizing the rendered appearance across the entire~\lod~chain~\cite{Weier:2023:Neural}, which often sacrifices coherent internal data representation. For fibrous materials specifically, recent approaches have focused on down-sampling volumetric data~\cite{Zhao:2016:Downsampling,Loubet:2018:New} using exclusively Symmetric GGX microflake (SGGX) distributions~\cite{Heitz:2015:SGGX}.  This limitation prevents the use of advanced fiber scattering models~\cite{Khungurn:2016:Matching}, as the original directional data (fiber tangents, density) becomes inseparably mixed with scattering information at each voxel.
\new{Zhou and colleagues~\cite{Zhou:2025:Appearance} follow a similar approach to tackle the problem of LoD in big scenes using an Aggregated Bidirectional Scattering Distribution Function (ABSDF), aggregating different appearances for multiples objects in a single voxel.}
Such mixing restricts material editability and constrains rendering to SGGX-compatible shading models.

\new{A special case for LoD generation are scenes that include elements at both macroscopic and microscopic level.
Hybrid mesh–volumes methods have been proposed to tackle this problem~\cite{Loubet:2017:Hybrid}; however, a mixed representation introduces its own challenges for some rendering techniques and require special-case handling.
Vicini et al. ~\cite{Vicini:2021:Nonexponential} proposed a non-exponential transmittance model that handles this issue while relaying just in the usage of volumes.}

Neural methods have also proven to be useful in LoD generation. Nevertheless, they either focus in geometry explicit representations, such as Signed Distance Functions (SDFs)~\cite{Takikawa:2021:Neural} or fail to properly capture high-frequency appearances~\cite{Weier:2023:Neural}.

Our approach overcomes these limitations by maintaining a clear separation between geometric/directional data and material properties throughout the~\lod~hierarchy. We preserve orientation distribution, thereby enabling the use of various anisotropic scattering models while maintaining visual consistency across detail levels. This separation also facilitates material editing at any level of detail without requiring regeneration of the entire hierarchy. Furthermore, our method efficiently handles the extreme sparsity of microgeometry through a parallel GPU implementation that processes only occupied regions, enabling high-resolution representation with manageable memory requirements.

\section{Voxelization for data aggregation}

\begin{figure*}[ht]
    \centering
    \includegraphics[width=1.0\textwidth]{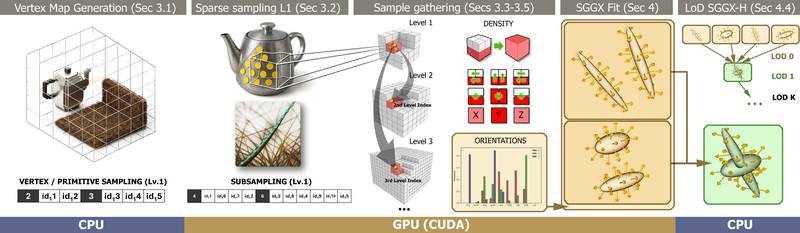}
    \caption{Our pipeline process from sparse 3D vertices to LoD of voxels in representing 3D generated model. We generate new \textit{vertices} subdividing our original primitives, each of these vertices with all information needed to be processes in the GPU\@. Our 3rd step gathers all valid samples already arranged in the corresponding voxel. Last, we process each generated histogram per voxel (for orientation data) and generate our whole hierarchical structure for rendering.}
    \label{fig:pipeline}
\end{figure*}

Traditional voxelization algorithms (Section~\ref{ssec:voxelization-methods}) require multiple passes for the generation of LoD structures, specially for geometries with high variability in local orientations, or surfaces with microscale data (e.g.: tangent maps for anisotropic reflections). Aggregating efficiently fine sparse details into coarser blocks is a complex task that usually implies rasterization and fixed-size memory allocation for sub-block grids. 

In this section, we explain our voxelization model, suitable for high frequency sparse geometric objects such as fibers, hair or thin planes with texture data, and designed to process volumetric data into multiple Levels of Detail (\lod), preserving as much appearance information as possible.

The voxelization pipeline is shown in Figure \ref{fig:pipeline}. Our model first samples the input 3D primitives, generating voxels only if they are filled by, at least one sample, thus avoiding unnecessary computation for non-occupied voxels. For each sample, we store as much data as necessary for future rendering purposes (Section \ref{sec:map_generation}). The generation of samples is different for curves and surfaces (Section \ref{sec:sparse_sampling}), but the data per node is stored and indexed in the same manner. Once the samples are generated at the maximum sampling rate and detail desired and stored as leaf nodes, they are gathered and aggregated into coarser bounding voxel blocks, in a multilevel hierarchical structure that is then used for \lod~generation. In our examples, we will use a three-level multi-grid, with resolutions optimized for each scenario and limited automatically by the samples generated at the geometry and the GPU memory available, but additional levels can be considered depending on the complexity of the scene. This gathering is computed in parallel on the GPU, composing histograms of orientation data (Section \ref{sec:directional_data}) and optical density for densities (Section \ref{sec:voxelization:density}) from the data stored at the leaf nodes. Furthermore, since our memory consumption does not depend of the grid resolution, it is specially suitable for high resolution voxelization of sparse, detailed geometry with a low percentage of volume occupancy: thin surfaces, hair, fibers, etc.

In Section~\ref{sec:postprocess}, we propose a novel LoD hierarchical representation based on SGGX for rendering, that leverages the data gathered at this stage. The results are shown in Section \ref{sec:results}.

\subsection{Vertex map generation}
\label{sec:map_generation}
The first stage of our algorithm is executed once per input model (triangle mesh or spline collection in our examples~\ref{fig:pipeline}).

In this step, we generate a map to keep the relationship between each sample of the primitive that we receive as input (triangle or spline), and the voxelization nodes sharing same space. 
We store multiple indices per node, referencing the geometric and material properties of the object at that point, and the multilevel blocks where the node is spatially registered. In practice, we can map this identifier to different materials (BSDF or Phase Function), for rendering purposes, or even parametric data to use in additional voxelization sub-steps to expand the original sample in the node to finer voxel resolutions. For instance, we can store the nodes of a spline representing the central axis of a yarn and a cross section displaying the distribution of secondary fibers at each node.

For each input model, we compute a global bounding box ($B(b_{min}, b_{max})$) and maximum generation radius $\delta$ (taking into account factors like maximum triangle edge length, radius, or displacement distance). Then, we move node data to GPU memory, where we use the bounding box information to transform our samples from global space to normalized coordinates in range $[0,  Resolution)$ based in our input resolution for first and second levels, and also our selected resolution for the third level (8 voxels per axis in the examples of this paper):

\begin{align}
    &Resolution = \prod_{i=1}^{3} Resolution_i \\ 
    &\mathbf{P}_{normalized} = \frac{\mathbf{P} - b_{min}}{b_{max} - b_{min}} \\
    &\mathbf{P}_{volume} = \mathbf{P}_{normalized} * Resolution 
\end{align}

After this step, each node holds the following information:

\begin{itemize} 
	\item \textbf{Translation}: The position of the node represented by three floating-point values.
    \begin{equation} %
    p = (x, y, z) \in \mathbb{R}^3
	\end{equation} %
    \item \textbf{Normal}: The main direction of crimp offset displacement (splines) or face normal (triangles) represented by three floating-point values.
    \begin{equation} %
    n = (n_x, n_y, n_z) \in \mathbb{R}^3 , \quad \| \text{n} \| = 1
	\end{equation} %
    \item \textbf{Tangent}: The direction to the following node (splines) or face tangent (triangles) represented by three floating-point values.
    \begin{equation} %
    t = (t_x, t_y, t_z) \in \mathbb{R}^3, \quad \| \text{t} \| = 1
	\end{equation} %
    \item \textbf{Properties ID}: An identifier for the corresponding properties, represented by a 32-bit integer: $id \in {Z}_{32}$
\end{itemize}

\noindent Once the data is uploaded to the GPU in our desired coordinate space, we run a parallel kernel to generate the maximum number of nodes by first level block. During this process, we can also identify (and discard) the blocks containing no data (invalid blocks). For this, we check each control node using a distance function:
\begin{align}
    &\mathbf{P}_{block} = S_{block} * I_{block} \\
    &C_{block} = \mathbf{P}_{block} + \frac{S_{block}}{2} \\
    &\mathbf{P}_{relative} = \left\| \mathbf{P}_{volume} - C_{block} \right\| \\
    &\mathbf{Q} = P_{relative} - \frac{S_{block}}{2} \\
    &\mathbf{D} = \left\| \max(\mathbf{Q}, \mathbf{0}) \right\|^2 + \min\left( \max(\mathbf{Q}_x,\mathbf{Q}_y, \mathbf{Q}_z), 0 \right)
\end{align}
\noindent where we can use block size ($S_{block}$) and 3D index ($I_{block}$) to compute the node position relative to a given block $\mathbf{P}_{relative}$ based on the position of the block itself $\mathbf{P}_{block}$. $\mathbf{P}_{relative}$ can then be used to compute the distance $D$ to consider nodes that, even if they fall outside of a block, could still generate new samples that fall within it (e.g., for fiber generation or triangle point generation). Therefore, we consider any node that ensures $D < \delta^2$, where $\delta$ is the maximum distance where data can be generated (e.g: maximum yarn radius, maximum triangle edge length). Using this equation, we can discard those nodes with distance lower than zero.

Once we know the maximum number of nodes per block (level 1), we use that information to initialize a CPU flat array with space for this number per block and an extra counter (Figure~\ref{fig:pipeline} left). Then, we use a kernel to write for each block the number of nodes that contain data (valid nodes) and the ID of the nodes.
At this point, we copy this mask to CPU to know which first level blocks are empty and generate an axis--aligned stencil, using the shape of the filled blocks for later usage during volumetric path tracing.

\subsection{Sparse data sampling}
\label{sec:sparse_sampling}

After computing the array of valid nodes per block, we can proceed to generate our voxel data. For this we run, only for non empty blocks, three different kernels that will do the following:

\begin{enumerate}
	\item Prepare initial nodes to interpolate.
	\item Subsample the data.
	\item Generate voxel nodes from new samples
\end{enumerate}

The first step prepares nodes for easier computation of later steps. This step may reorder, duplicate or generate additional data so later is easier to access consecutive (connected) samples.  Our initial nodes may not be enough to cover all space continuously at the desired voxelization resolution. To avoid discontinuities we introduce a sub-sampling step. Notice that, for higher output resolutions, a higher number of sub-samples will be required if the input is really sparse. The last step generates samples from the interpolated new vertices and stores them in a given voxel based on their 3D position. Additionally, this step can be used to further increase detail of the samples, adding texture based data (3D or 2D) to modify properties of the nodes before storing them in a given voxel (eg: Normals, tangents, twist).

Per sample, we compute: a second and third level indices, the node orientation (tangent and normal data) and a single material index we can use to later apply different spatially varying properties such as Phase-functions or BSDFs for each node  (Table~\ref{tab:node_information}).

Even though all steps may differ somewhat for each input primitive, the second step is particularly dependent on it.
As an example, for splines defined using \textit{Catmull-Rom} interpolation~\cite{Catmull:1974:Splines} we want to first define the whole spline from the 4 required points that we store in memory, and then generate additional samples along the shape of the spline using linear spline interpolation.
In case we use triangles as our input primitive, our first step is easier since each triangle is defined in our data-model by points that are not shared between different triangles.
During the second step, we also generate additional points in the surface of each triangle. We use barycentric coordinates for this sampling.

Since all of these steps are computed in the GPU, relying on actual random sampling is not feasible. Instead, we use equally distributed samples based in kernel indices. We distribute the amount of samples chosen as input along $t$ for later usual interpolation when voxelizing spline primitives, and the low--distortion mapping proposed by Heitz~\cite{Heitz:2019:Triangle-map}, modified for GPU to avoid branching. Additionally, we weight the amount of final samples to generate in each single triangle based on their relative area to the biggest triangle in the model. We show a comparison between the Naive method and this approach in the Supplementary Material.

\subsection{Sample gathering}
\label{ssec:sample-gathering}
Once the samples are generated, we move this data to a grid-like structure, as done by previous methods~\cite{Museth:2013:VDB,Museth:2021:nanovbd}. By saving just the last sample generated at each voxel, in an asynchronous fashion, we avoid the implicit ordering of draw calls that previous algorithms suffer (multiple calls per axis) since we can order the nodes in the GPU. Such ordering also facilitates data aggregation, histogram computation, and further processing.

Once our samples are ordered by second level block ID, we use a kernel to detect the vector positions where the ID changes, so we know where the samples for a given block start and how many of them we computed. Note that, since samples are the main data, the amount of memory that we need to prepare for a worst-case scenario depends on the amount of samples generated instead of the voxel resolution. Even if the resolution scales dramatically, we do not waste any memory in empty voxels. 

\subsection{Directional data}
\label{sec:directional_data}

Thanks to the per-voxel sample gathering described in Section~\ref{ssec:sample-gathering}, we can compute a distribution histogram for any directional data, such as normals or tangents, at every voxel with $density > 0$.

While this histogram gives us a higher detail information for the underlying data inside a voxel than a single direction, using this information for rendering can be difficult. Aside from the space needed to store these histograms for each voxel, sampling a direction from it (needed if we want to use the volume for volumetric rendering) is also computationally expensive.

To overcome this, we propose to fit each computed histogram to a SGGX function \cite{Heitz:2015:SGGX}, which is considered a representation with a good trade-off between compression and quality for 3D orientation distributions. Since SGGXs represent a spheroid--like shape, and can be easily sampled, we can use them to encode the distribution of directions in the histogram, using just 6 values. Furthermore, they are easily aggregated into new SGGX distributions, which we leverage in the algorithm presented in Section~\ref{sec:postprocess}.

\subsection{Density occupation and view dependent data}
\begin{table}[]
	\centering
	\begin{tabular}{ccc}
		\toprule
		Parameters                          & Type       & Bytes \\
		\hline                      
		\multicolumn{1}{l}{2nd level index} & integer    & 4     \\
		\multicolumn{1}{l}{3rd level index} & integer    & 4     \\
		\multicolumn{1}{l}{Tangent data}    & u--integer & 3x1   \\
		\multicolumn{1}{l}{Normal data}     & u--integer & 3x1   \\
		\multicolumn{1}{l}{Material ID}     & u--integer & 4     \\

		\bottomrule
	\end{tabular}

	\caption[Voxelization by insertion]{Required space for our initial information per node. This data is used to compute the information that we stored in the final leaf nodes (in our examples, the voxels at the 2nd resolution level).}
	\label{tab:node_information}
\end{table}

\label{sec:voxelization:density}

Many real-world materials can only be accurately represented using correlated media~\cite{Bitterli:2018:Radiative, Guo:2019:Fractional}. Thus, storing a single value for density data without any consideration for the view direction, usually produces inaccurate rendering results.

As a solution, we compute the probability of an event depending of the view direction, using the finer resolution level in the voxelization (level 3), in a similar fashion to Crassin et al. \cite{Crassin:2011:Interactive} we project the 3D voxel block grid to a 2D grid. Samples are projected in the $XY$, $XZ$ and $YZ$ planes using GPU shared memory, and we store the projected position of every sample by block, so we can obtain a value per axis, producing an anisotropic voxel (Figure~\ref{fig:pipeline}).

To compute occupancy $\mathbf{O}$, instead of projecting the data to the axis, we analyze the third level data, obtaining a percentage by dividing the number of different identifiers by the amount of voxels in our third level of voxelization, he   nce computing actual density for any given voxel of second level:

\begin{align}
    \mathbf{O} = \frac{\sum_{v \in \text{block3}} \text{hit}(v)}{Resolution_3^3}
\end{align}

\section{Data post-processing and fitting}
\label{sec:postprocess}

Our voxelization process ends with different samples per voxel, that we are able to use to compute an histogram of orientations. To generate the final volume, we could store randomly one of them as the final value, hence arriving at the same data that can be generated with classic voxelization approaches.
However, we can use all this samples per voxel to perform an aggregation step and arrive to a better representation for voxel values compared to a single random value/sample, also easier to handle than RAW histograms.

For our pipeline, we focus in fitting this data using a single SGGX distribution~\cite{Heitz:2015:SGGX} at leaf-voxel level. Our approach could be used to fit more complex distribution instead, or even perform sampling using the raw data from the histogram.
We use the original distribution presented by Heitz and colleagues without its phase-function interpretation, since our only requirement is to store directional data such as tangents and directional density information for later evaluation, with a different phase function model. By just storing orientation directly by using SGGX instead, we are free to use any material model later, using the frame defined by our sampled orientation data during render.

\subsection{Fitting to SGGX distribution}\label{sec:fitting}

To generate our single distribution for the leaf-level voxel we use the approach proposed by Heitz~\cite{Heitz:2015:SGGX} to fit arbitrary distributions to their SGGX distribution.

\noindent We compute the covariance matrix $\mathcal{E}$ for our orientation data $X$:

\begin{align}
    X = \{\mathbf{x}_1, \mathbf{x}_2, \dots, \mathbf{x}_n\},    \mathbf{x}_i \in \mathbb{R}^3\\
    \mathcal{E} = \frac{1}{n} \sum_{i=1}^{n} (\mathbf{x}_i - \bar{\mathbf{x}})(\mathbf{x}_i - \bar{\mathbf{x}})^\top
\end{align}

From this covariance matrix we can later extract eigenvector $(\omega_1, \omega_2, \omega_3)$ and their corresponding eigenvalues $\{\lambda_1, \lambda_2, \lambda_3\}$, equivalent to the projected areas on each eigenvector direction~\cite{Heitz:2015:SGGX}:
\begin{equation}
    \lambda_1 = \sigma(\boldsymbol{\omega}_1), \quad \lambda_2 = \sigma(\boldsymbol{\omega}_2), \quad \lambda_3 = \sigma(\boldsymbol{\omega}_3)
\end{equation}
Using this computed values we can then get the SGGX matrix $S$ defining the distribution of the original histogram as:
\begin{equation}
    \small
	S = (\omega_1, \omega_2, \omega_3)
	\begin{pmatrix}
		\sigma(\omega_1) & 0                & 0                \\
		0                & \sigma(\omega_2) & 0                \\
		0                & 0                & \sigma(\omega_3)
	\end{pmatrix}
	(\omega_1, \omega_2, \omega_3)^T
	\label{eq:SGGX-from-fit}
\end{equation}
We can sample this SGGX distribution to obtain a direction from the inner orientation distribution allowing scattering functions that rely in tangential/normal orientation to be used for rendering. Different approaches such as microflakes/microfacets models can be used too, re-oriented using the distribution stored at each voxel.

Our sampling process uses a similar approach as the one presented by Heitz et al.~\cite{Heitz:2015:SGGX} for the Visible Normal Distribution Function (VNDF). We modify their approach by generating uniform samples in the whole sphere, instead of the hemisphere, so our samples can be projected from the whole space to the underlying ellipsoid. This ensures that our samples get distributed in the whole Normal Distribution Function (NDF), not only in the VNDF subsection. When sampling SGGX distributions for rendering, we use the VDNF as usual. We also modify our fitting step to account for corner cases where all the input orientations align to almost the exact same orientation, hence, generating a ill-defined SGGX. We use a uniform distribution to introduce a small deviation to the original data so we obtain a noisier but unbiased SGGX as output.

\subsection{Density data}
\label{sec:processing:density}

For density data, it is not trivial to encode directional occupation using SGGX distributions, since occupation does not fit a set of main axes, and requires multiple evaluations of a given number of distributions that will increase exponentially with the LoD selected. Moreover, SGGX distributions are not normalized and the normalization term has a non-closed form. While this lack of normalization does not affect orientation encoding, evaluating the area of the encoded ellipsoid for density storage, requires this normalization factor, or at least the scale of the ellipsoid volume. 

Taking into account these limitations, we compute LoD density levels with the same projection approach described in Section~\ref{sec:voxelization:density}. We project nodes at the finer resolution level to the $XY$, $XZ$ and $YZ$ planes, obtaining a value per axis for the coarser level.

\subsection{Storing data}

Once we have computed the SGGX orientation distributions, and LoD densities, we need to store them in the final volumes that will be used for volumetric rendering.
Using GPU kernels, we store the main indices of the SGGX matrix representing orientation data ($S$) in the leaf nodes of the LoD pyramid (LoD 0 voxels), using the compact storage presented in the original work (Equation~\ref{eq:compact-sggx}) that expresses the parameters in linear space, allowing the use of 1 byte per parameter, so each distribution takes 6 bytes of space.

\begin{equation}
	\begin{aligned}
		\sigma_x & = \sqrt{S_{xx}},                     & \sigma_y & = \sqrt{S_{yy}},                     & \sigma_z & = \sqrt{S_{zz}},                    \\
		r_{xy}   & = \frac{S{xy}}{\sqrt{S_{xx}S_{yy}}}, & r_{xz}   & = \frac{S{xz}}{\sqrt{S_{xx}S_{zz}}}, & r_{yz}   & = \frac{S{yz}}{\sqrt{S_{yy}S_{zz}}}
		\label{eq:compact-sggx}
	\end{aligned}
\end{equation}

For higher (coarser) levels we can compute new volumes of larger voxel blocks using SGGX interpolation~\cite{Heitz:2015:SGGX} to aggregate voxels in a $2x2x2$ structure. However, joining all distributions into a single new SGGX distribution produces a loss of accuracy, specially with highly anisotropic distributions. This inaccuracy is increased with each interpolation, producing very isotropic results, that will yield rougher scattering at the rendering stage (See the naive SGGX results in our Supplementary Material).

For better quality, we can use more than a single SGGX distribution to represent this anisotropy, by referencing lower level distributions and using them together to sample the mixture of orientations. However, such approach increases exponentially the number of memory operations and samples in rendering time for coarse levels. Thus, we propose an aggregation method on top of the previous approach, merging similar distributions so we control the storage needed to a maximum amount of distributions per LoD block.

Our method retains the most relevant distributions, even if they encode perpendicular orientations. In our experiments, we obtained good results with a maximum of three (3) distributions per level, which retain information in three main axes, even if the data we want to fit is evenly distributed. For density and occupancy, we store the values computed as explained in Section~\ref{sec:voxelization:density}.

\subsection{LoD computation}
\label{sec:lod-sggx}
Accurately representing a complex mix of orientations may require more SGGX functions than we expect to store in a volume. 
A simple workaround can be designed to reduce their number to a single, representative, SGGX, by sampling multiple times each original SGGX, concatenate their samples and generate the end SGGX using the fitting algorithm described in Section~\ref{sec:fitting}. However, as discussed previously, these distributions tend to be highly anisotropic, and such sampling produces a drastic information loss.

To overcome these problems, and to allow for a more fine-grained LoD calculation, we propose a simple aggregation method that progressively reduces the number of SGGXs used to represent the needed directional data, whilst preserving more high-frequency details than what could be accomplished using a simpler aggregation method. We draw inspiration from classical unsupervised learning algorithms in machine learning and pose this problem as a \textit{hierarchical clustering} method~\cite{Murtagh:2012:Algorithms}. In particular, we build a \textit{dendrogram} of LoDs, by progressively reducing the original set of distributions, as we illustrate on rightmost part of Figure~\ref{fig:pipeline}. Given a set of distributions $D_L = \{S_1, S_2, \ldots, S_L\}$ for LoD $L$, we compute the next LoD $L-1$ simply by aggregating the two most similar SGGXs at $L$. This aggregation method preserves all the information present in the rest of SGGX distributions and loses the least amount of information, as the two most redundant SGGX are substituted by a single, representative one.

Instead of computing the similarity between two SGGX distributions by comparing their defining parameters (see equation~\ref{eq:compact-sggx}), we compare the histograms of their samples. This way, we compute distance between SGGXs in terms of the underlying probability distributions they encode.  More specifically, for each SGGX $S_i$ at each $L$, we draw $N$ samples, which we bin into a histogram $H(S_i)$. Using a probability distribution distance function $dist(d_1,d_2)$, we then compute the distance between each pair of SGGX at $L$. We obtain the two most similar distributions searching for the smallest distance between each pair of histograms, as described on equation~\ref{eq:similarity}:

\begin{equation}\label{eq:similarity}
    \argmin\limits_{i,j}  \ dist(H(S_i), H(S_j)), \quad where \ i\neq j; S_i, S_j \in D_L
\end{equation}

\allowbreak
As we use a symmetric distance function ($dist(d_1,d_2)=dist(d_2,d_1)$), this distance can be computed only once for each pair of distributions (eg $\forall i > j; S_i, S_j \in D_L $), resulting in a strictly triangular matrix of distances, for which this minimum is more easily found. We then compute a new distribution by aggregating the samples of $S_i$ and $S_j$, and fit their resulting histogram into a new SGGX $S_n$. The next LoD $D_{L-1}$ will simply be $D_L$ without $S_i$ and $S_j$ and with the new $S_n$, as in:  $D_{L-1} = (D_L\setminus\{S_i,S_j\}) \cup S_n $. Running this algorithm until $L=1$, we can build a hierarchical aggregation of LoDs, which provides fine-grained control over the trade-off between computational efficiency and information preservation.

\textbf{Implementation details}: We use $N=5000$ samples for each $S$, and compute their histograms $H(S)$ using $5$ bins in each spatial dimension. For comparing the histograms, we use the x\textit{Wasserstein} distance, which is widely used in probability and machine learning algorithm~\cite{Vallender:1974:Wasserstein, Heitz:2021:Wasserstein, Gulrajani:2017:Improved}.

\section{Results}
In this section, we show the capabilities of our method, in terms of voxelization computational cost, rendering quality, and quantitative comparisons with baselines.

\newcommand{\ImageComparison}[1]{
    \includegraphics[width=\textwidth, height=.5625\textwidth]{#1}
}

\newcommand{\comparisonGeneral}[3]{
  & \ifthenelse{\equal{#1}{}}{\includegraphics[width=\subFigW, height=\subFigH,keepaspectratio]{example-image-duck}}{\includegraphics[width=\subFigW, trim={0cm 0cm 0cm 0cm}, clip]{#1}} 
  & \ifthenelse{\equal{#2}{}}{\includegraphics[width=\subFigW, height=\subFigH,keepaspectratio]{example-image-duck}}{\includegraphics[width=\subFigW, trim={0cm 0cm 0cm 0cm}, clip]{#2}} 
  & \ifthenelse{\equal{#3}{}}{\includegraphics[width=\subFigW, height=\subFigH,keepaspectratio]{example-image-duck}}{\includegraphics[width=\subFigW, trim={0cm 0cm 0cm 0cm}, clip]{#3}}
}

\begin{figure}[ht]
    \newcommand\subFigW{0.29\columnwidth}
    \newcommand\subFigH{50px}
    \centering
    \begin{tabular}{
        >{\centering\arraybackslash}m{0.02\columnwidth}
        >{\centering\arraybackslash}m{0.28\columnwidth}
        >{\centering\arraybackslash}m{0.28\columnwidth}
        >{\centering\arraybackslash}m{0.28\columnwidth}}
         & \textbf{Ground Truth}                 
         & \textbf{Averages}                                                                      
         & \textbf{Ours} \\
        \midrule
        {\rotatebox{90}{\textbf{Hibiscus}}}      
         \comparisonGeneral{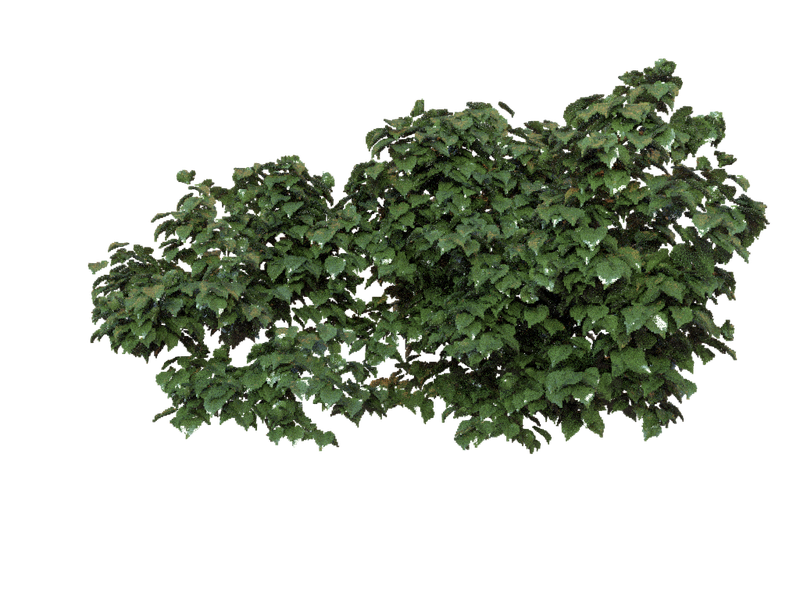}{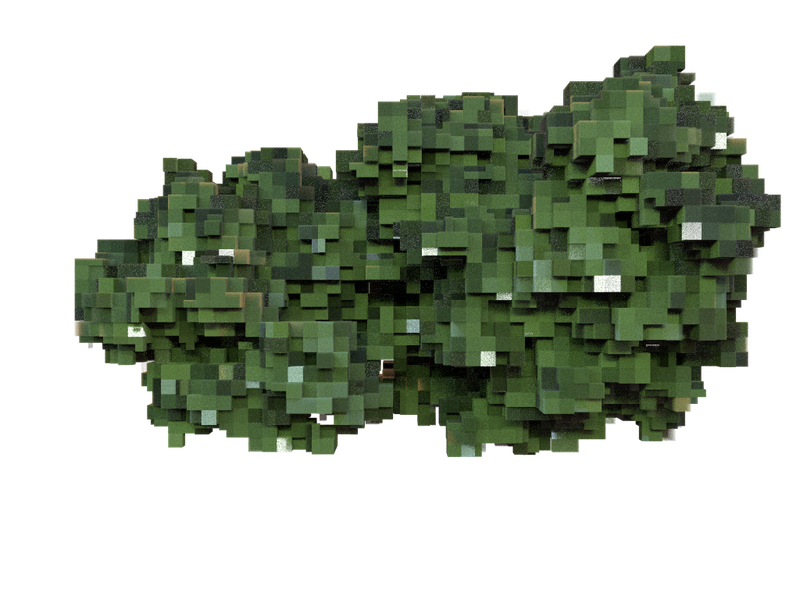}{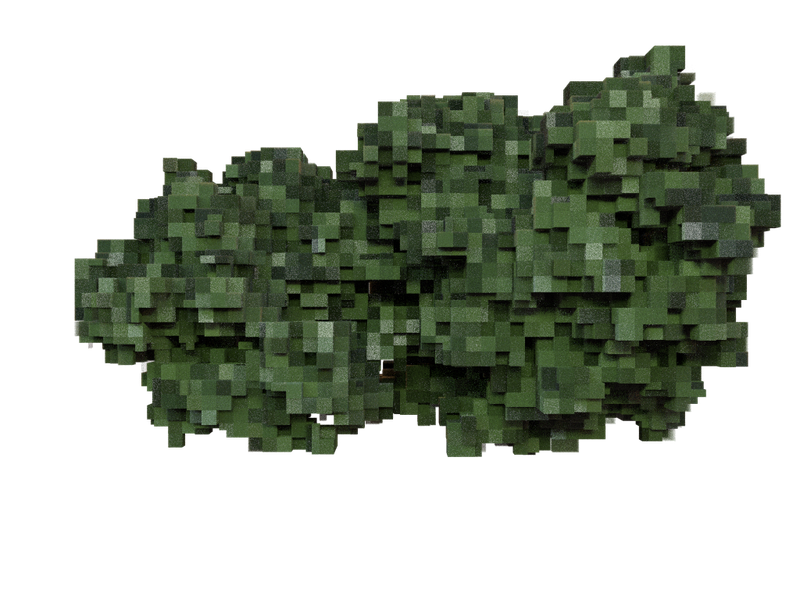}
         \\%
        {\rotatebox{90}{\textbf{Grass}}}      
         \comparisonGeneral{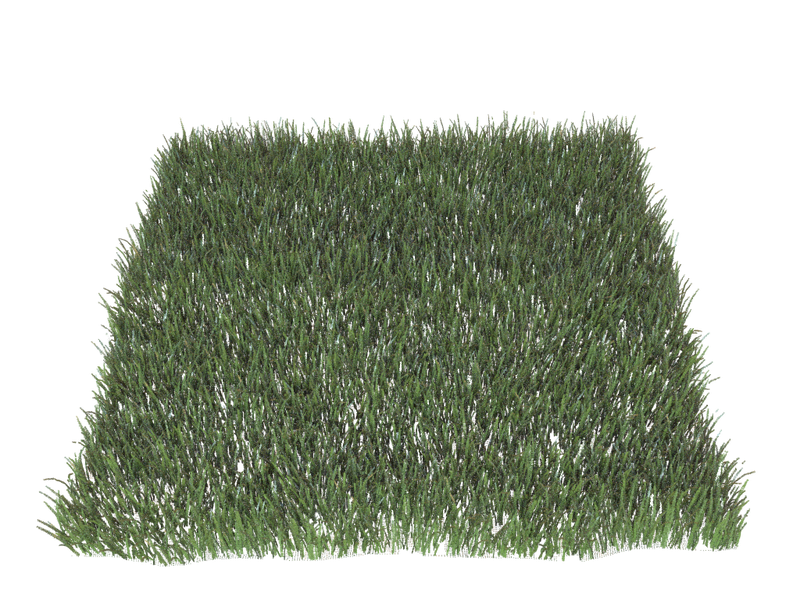}{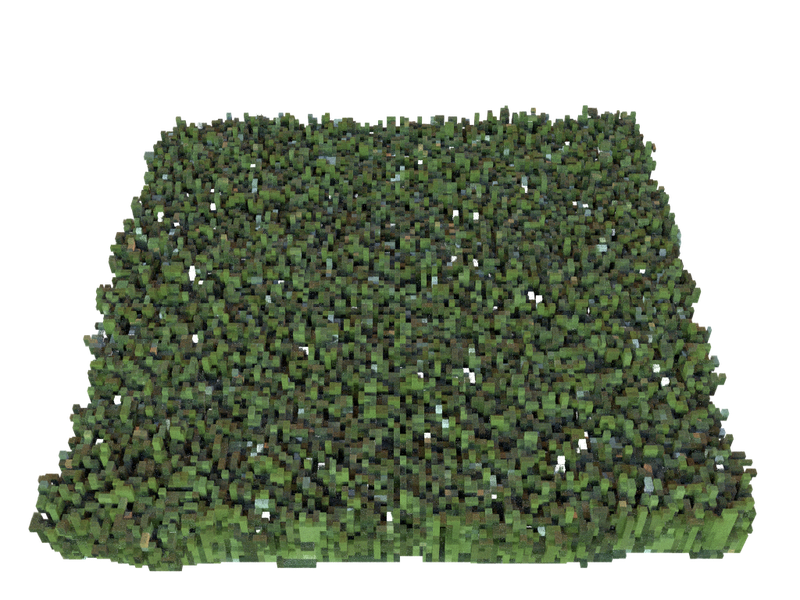}{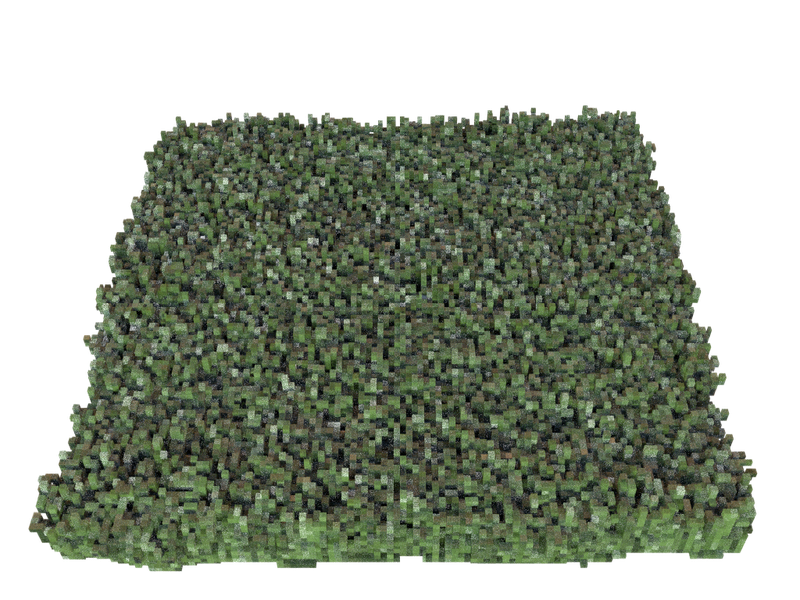}
         \\%
         {\rotatebox{90}{\textbf{Gardenia}}}      
         \comparisonGeneral{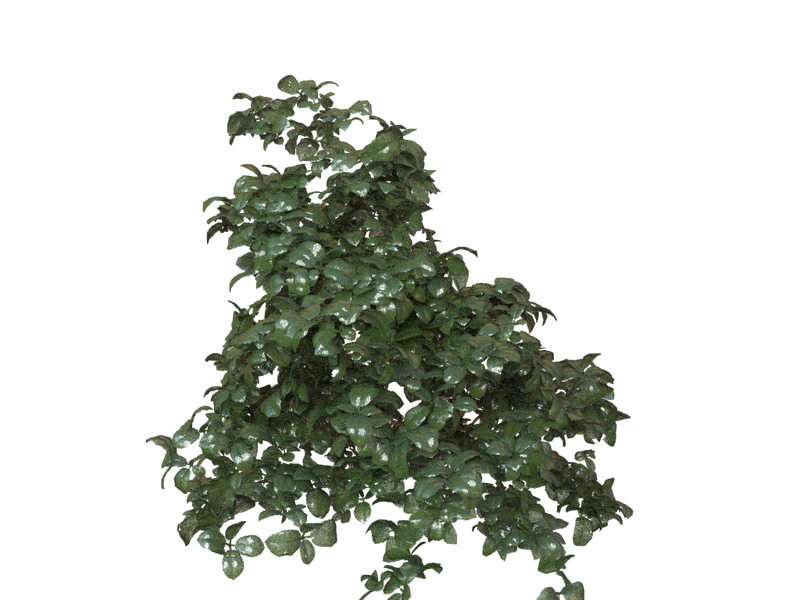}{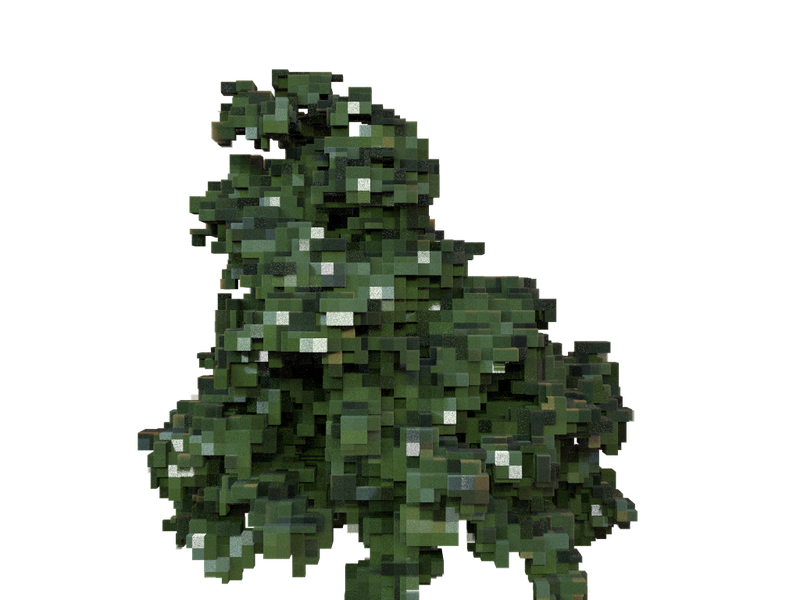}{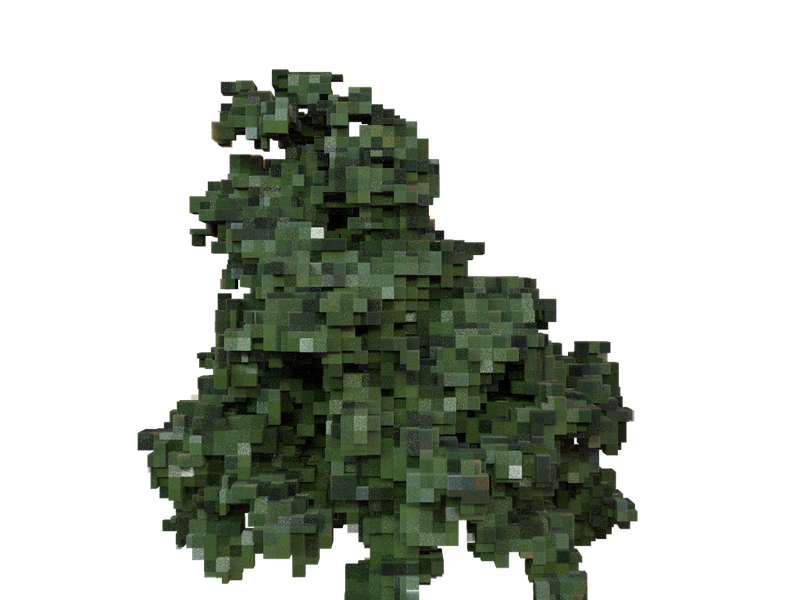}
    \end{tabular}
    \caption{Multiple comparisons for vegetation models from Hasselgen~\cite{Hasselgren:2021:Appearance} dataset (Hibiscus and Gardenia) and grass model from \href{https://free3d.com/}{free3d.com}. We provide resolution data and \FLIP metrics for these scenes in our Supplementary Material.}
    \label{fig:vegetation-comparisons}
\end{figure}

\newcommand{\comparisonhair}[3]{
  & \ifthenelse{\equal{#1}{}}{\includegraphics[width=\subFigW, height=\subFigH,keepaspectratio]{example-image-duck}}{\includegraphics[width=\subFigW, trim={10cm 5cm 10cm 0cm}, clip]{#1}} 
  & \ifthenelse{\equal{#2}{}}{\includegraphics[width=\subFigW, height=\subFigH,keepaspectratio]{example-image-duck}}{\includegraphics[width=\subFigW, trim={10cm 5cm 10cm 0cm}, clip]{#2}} 
  & \ifthenelse{\equal{#3}{}}{\includegraphics[width=\subFigW, height=\subFigH,keepaspectratio]{example-image-duck}}{\includegraphics[width=\subFigW, trim={10cm 5cm 10cm 0cm}, clip]{#3}}
}

\newcommand{\comparisonhairNatural}[3]{
  & \ifthenelse{\equal{#1}{}}{\includegraphics[width=\subFigW, height=\subFigH,keepaspectratio]{example-image-duck}}{\includegraphics[width=\subFigW, trim={5cm 0cm 9cm 2cm}, clip]{#1}} 
  & \ifthenelse{\equal{#2}{}}{\includegraphics[width=\subFigW, height=\subFigH,keepaspectratio]{example-image-duck}}{\includegraphics[width=\subFigW, trim={5cm 0cm 9cm 2cm}, clip]{#2}} 
  & \ifthenelse{\equal{#3}{}}{\includegraphics[width=\subFigW, height=\subFigH,keepaspectratio]{example-image-duck}}{\includegraphics[width=\subFigW, trim={5cm 0cm 9cm 2cm}, clip]{#3}}
}

\begin{figure}[ht]
    \newcommand\subFigW{0.29\columnwidth}
    \newcommand\subFigH{50px}
    \centering
    \begin{tabular}{
        >{\centering\arraybackslash}m{0.02\columnwidth}
        >{\centering\arraybackslash}m{0.28\columnwidth}
        >{\centering\arraybackslash}m{0.28\columnwidth}
        >{\centering\arraybackslash}m{0.28\columnwidth}}
         & \textbf{Ground Truth}                 
         & \textbf{Averages}                                                     
         & \textbf{Ours} \\
        \midrule
        {\rotatebox{90}{\textbf{Wavy}}}      
         \comparisonhair{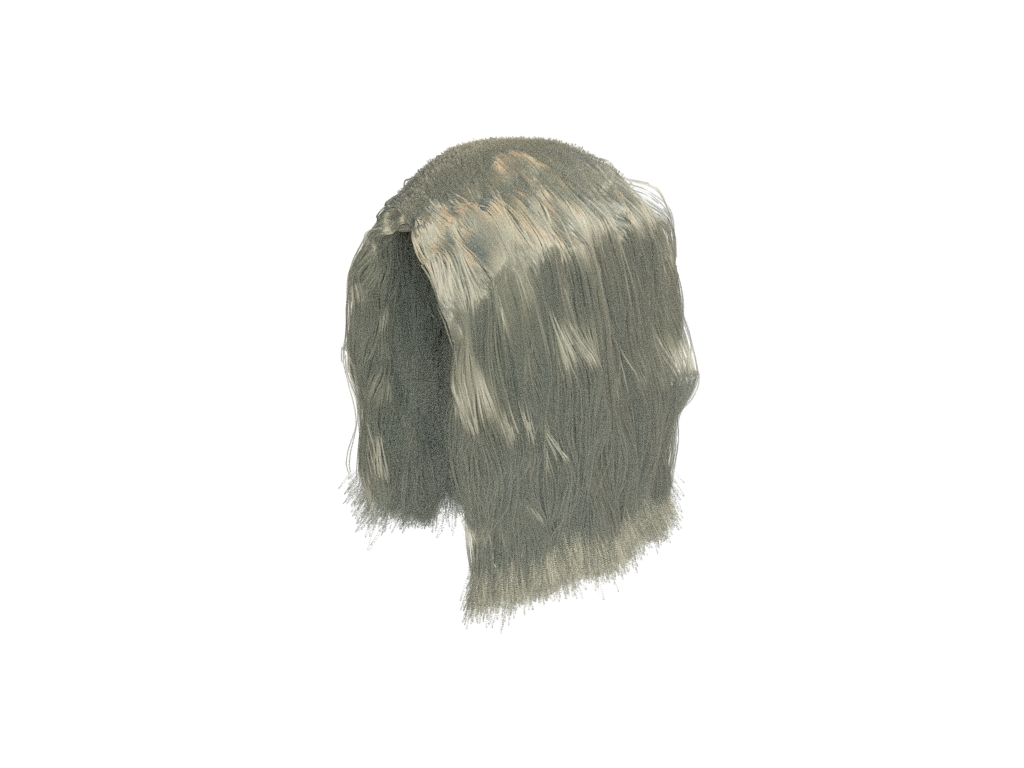}{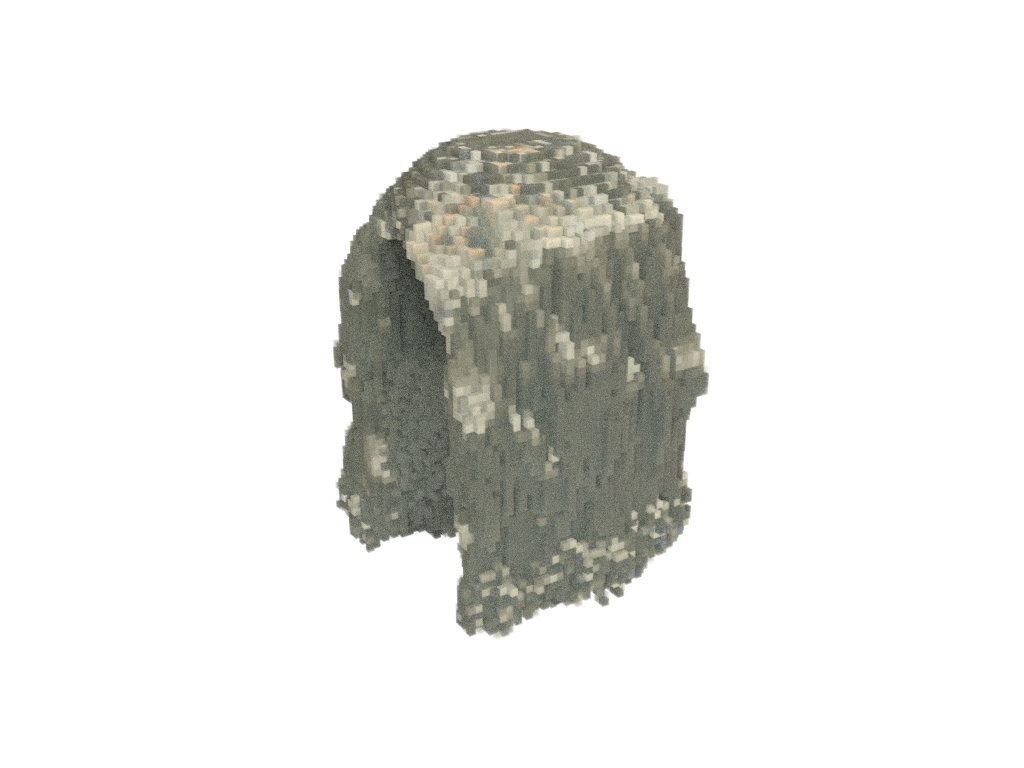}{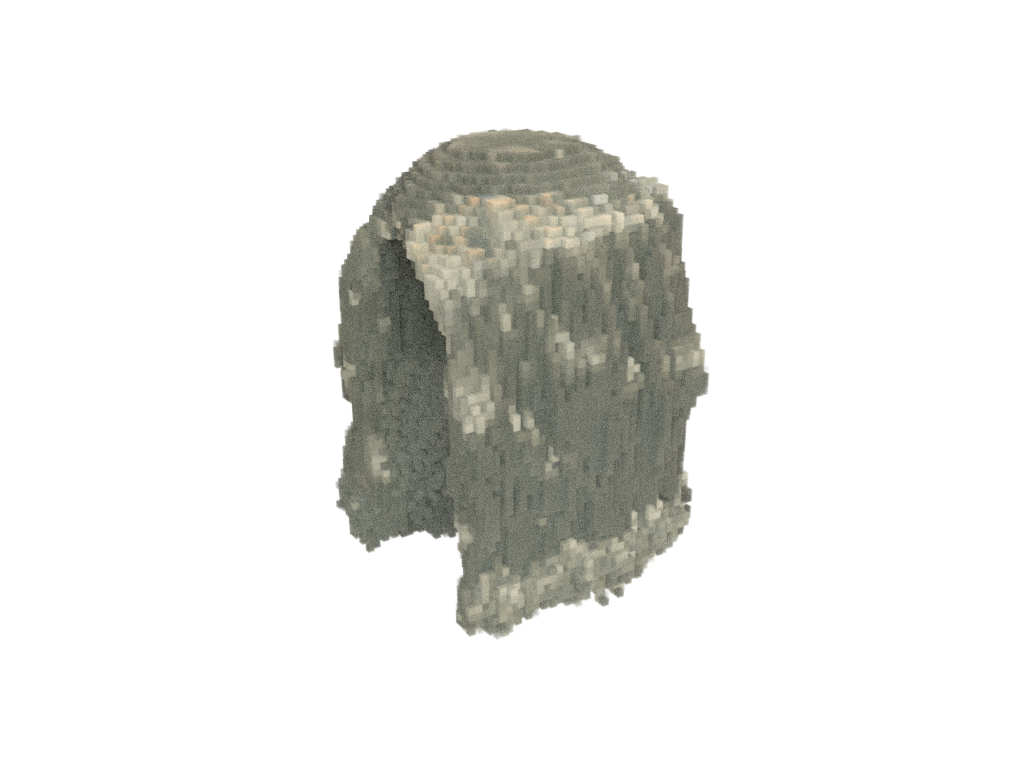}
         \\%
         {\rotatebox{90}{\textbf{Natural}}}      
         \comparisonhairNatural{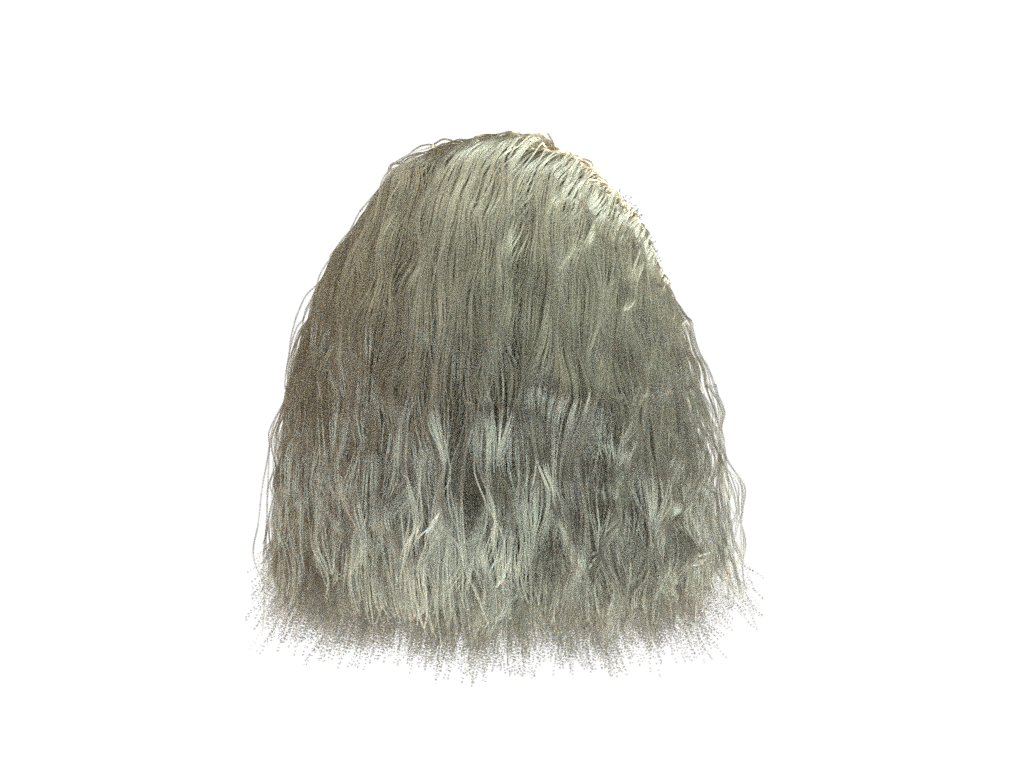}{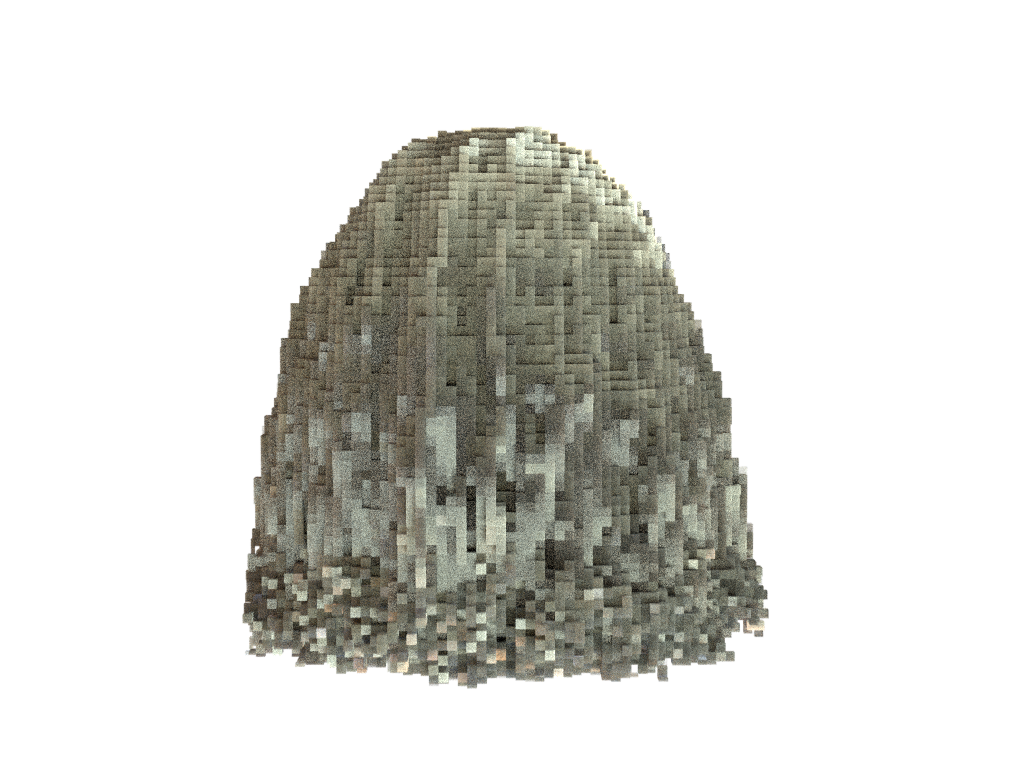}{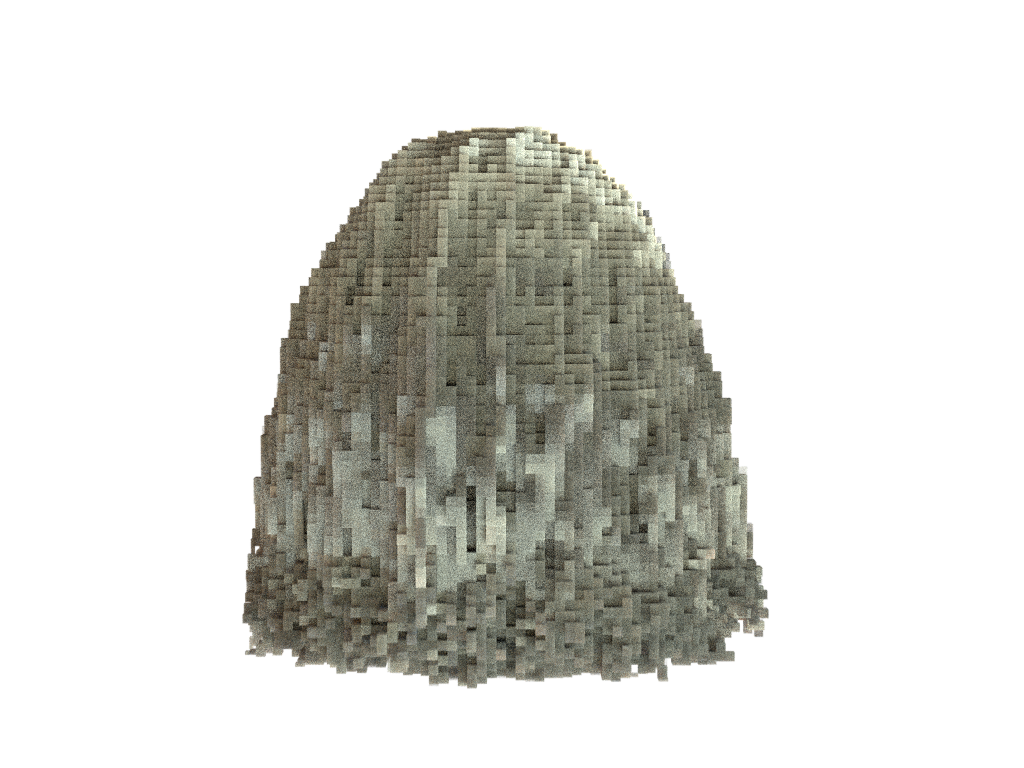}
         \\%
    \end{tabular}
    \caption{Comparisons of our Level of Detail method against Naive downsampling for Yuksel's hair model. We translated the original files to splines to fit our pipeline and then proceed to voxelize them to generate a Ground Truth render from the original voxelized volume. We provide resolution data and \FLIP metrics for these scenes in our Supplementary Material.}
    \label{fig:hair-comparisons}
\end{figure}

\begin{figure}[]
    \newcommand\subFigW{0.26\columnwidth}

    \centering
    \begin{tabular}{
        >{\centering\arraybackslash}m{0.02\columnwidth}
        >{\centering\arraybackslash}m{0.23\columnwidth}
        >{\centering\arraybackslash}m{0.23\columnwidth}
        >{\centering\arraybackslash}m{0.23\columnwidth}}
         & \textbf{Ground Truth}             
         & \textbf{Averages}                                       
         & \textbf{Ours}                                            
         \\
        \midrule
        \multirow{2}{*}[+6pt]{\makecell[c]{\rotatebox{90}{\textbf{LoD 1}}}}
         & \includegraphics[width=\subFigW,trim={0cm 0cm 0cm 0cm}, clip]{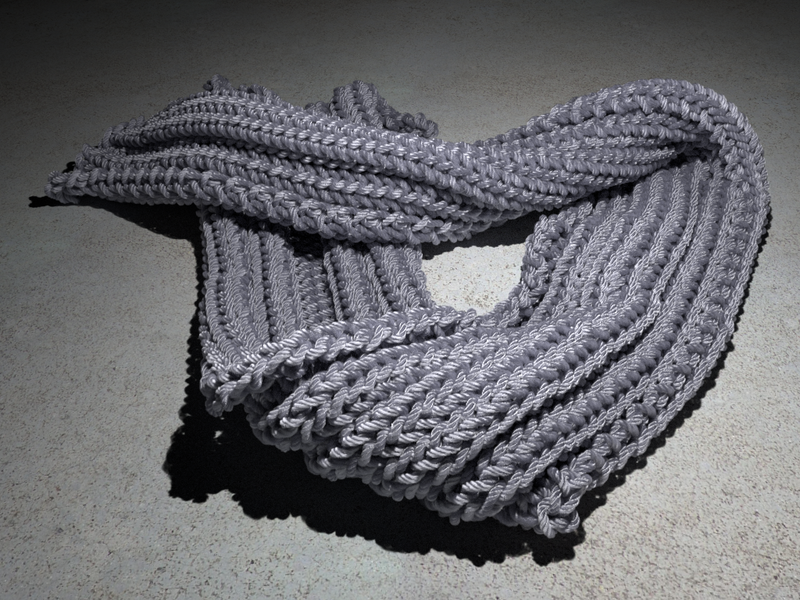}
         & \includegraphics[width=\subFigW,trim={0cm 0cm 0cm 0cm}, clip]{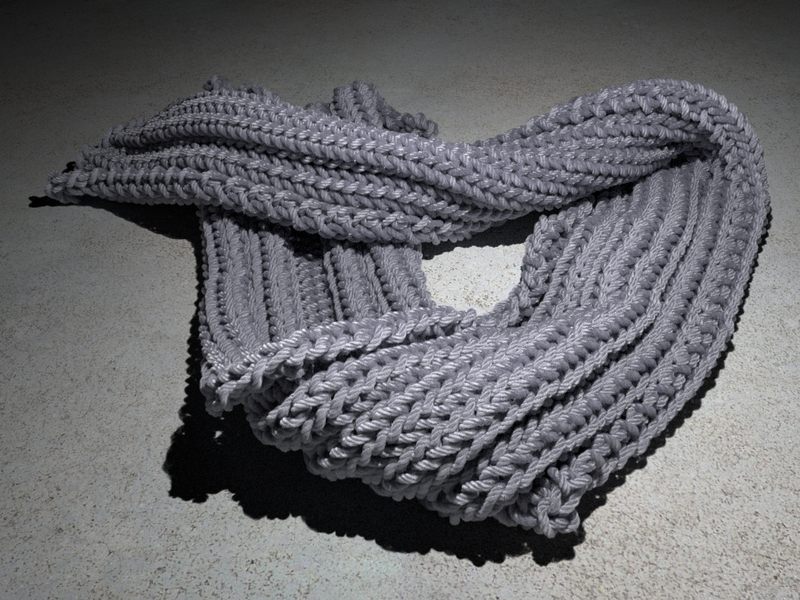}       
         & \includegraphics[width=\subFigW,trim={0cm 0cm 0cm 0cm}, clip]{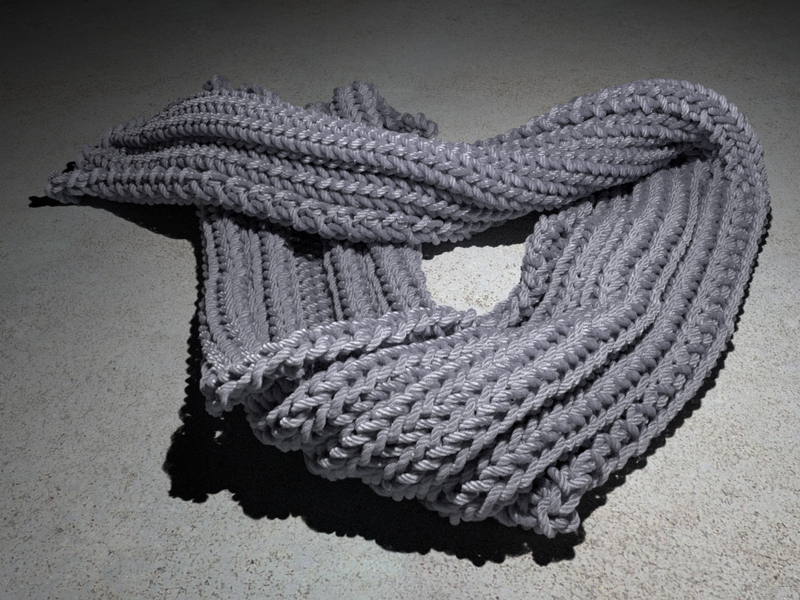}
         \\
        \midrule
        \multirow{2}{*}[+6pt]{\makecell[c]{\rotatebox{90}{\textbf{LoD 2}}}}
         & \includegraphics[width=\subFigW,trim={0cm 0cm 0cm 0cm}, clip]{figures/scarf/gt.png}
         & \includegraphics[width=\subFigW,trim={0cm 0cm 0cm 0cm}, clip]{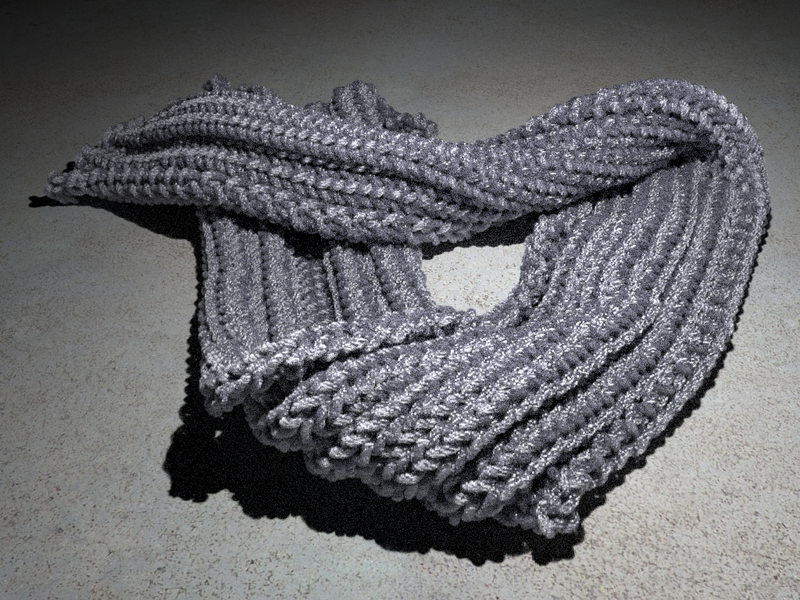}  
         & \includegraphics[width=\subFigW,trim={0cm 0cm 0cm 0cm}, clip]{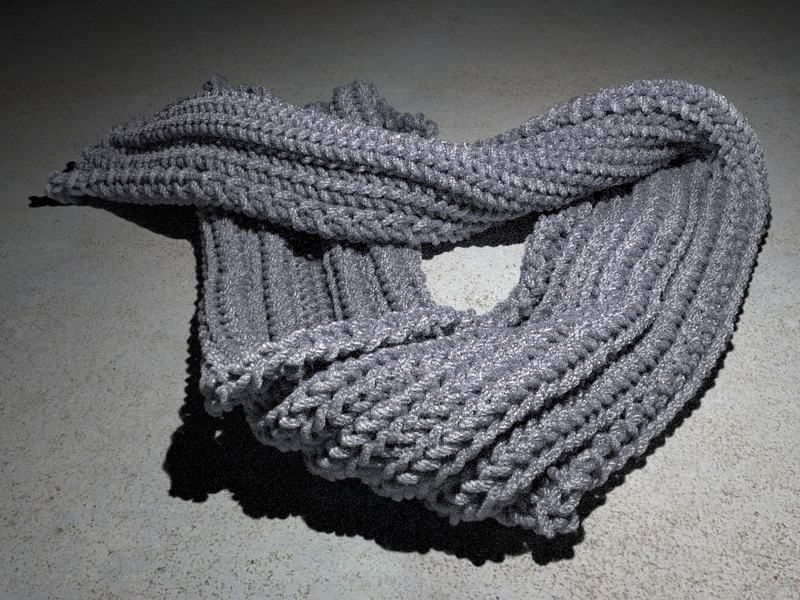}
         \\
        \midrule
        \multirow{2}{*}[+6pt]{\makecell[c]{\rotatebox{90}{\textbf{LoD 3}}}}
         & \includegraphics[width=\subFigW,trim={0cm 0cm 0cm 0cm}, clip]{figures/scarf/gt.png}
         & \includegraphics[width=\subFigW,trim={0cm 0cm 0cm 0cm}, clip]{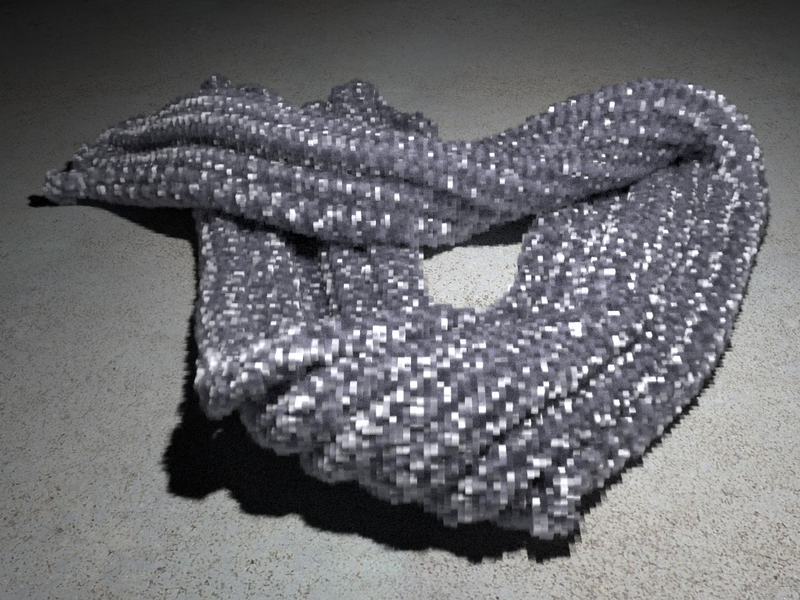}          
         & \includegraphics[width=\subFigW,trim={0cm 0cm 0cm 0cm}, clip]{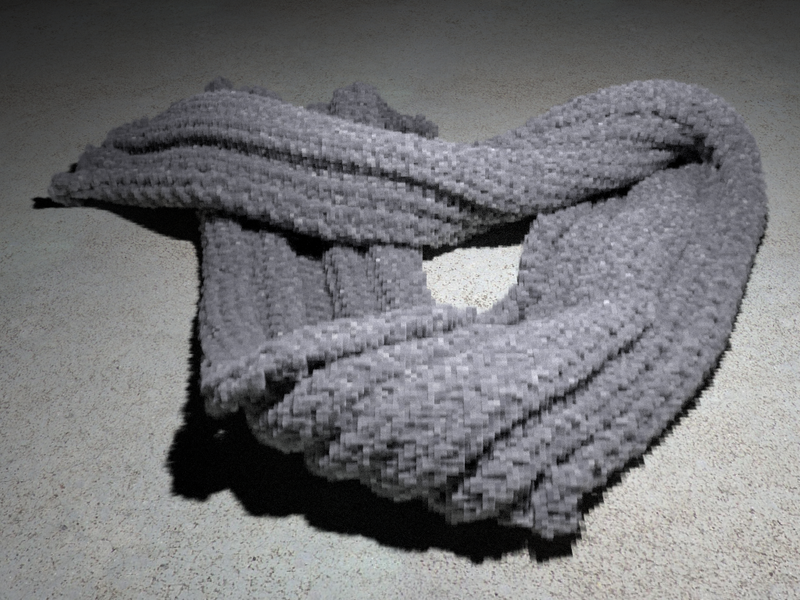}
         \\
        \midrule
    \end{tabular}
    \caption{Comparisons of renders at different voxelization LoDs of the Scarf volume~\cite{Jakob:2010:Radiative} using Ground Truth data (left) and our hierarchical model (right) for tangent data.}
    \label{fig:scarf-comparisons}
\end{figure}

\begin{figure}[]
    \newcommand\subFigW{0.26\columnwidth}

    \centering
    \begin{tabular}{
        >{\centering\arraybackslash}m{0.02\columnwidth}
        >{\centering\arraybackslash}m{0.23\columnwidth}
        >{\centering\arraybackslash}m{0.23\columnwidth}
        >{\centering\arraybackslash}m{0.23\columnwidth}}
         & \textbf{Ground Truth}             
         & \textbf{Averages}                                       
         & \textbf{Ours}                                            
         \\
        \midrule
        \multirow{2}{*}[+9pt]{\makecell[c]{\rotatebox{90}{\textbf{LoD1}}}}
         & \includegraphics[width=\subFigW,trim={1cm 0cm 1cm 3cm}, clip]{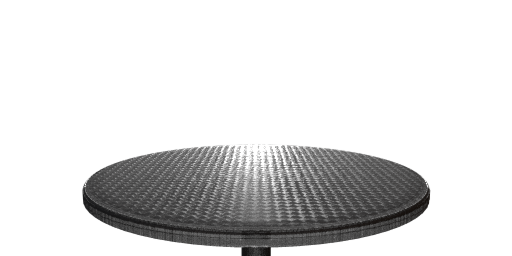}
         & \includegraphics[width=\subFigW,trim={1cm 0cm 1cm 3cm}, clip]{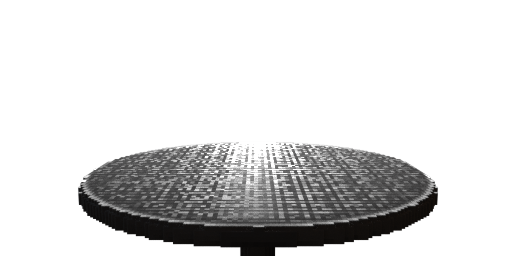}       
         & \includegraphics[width=\subFigW,trim={1cm 0cm 1cm 3cm}, clip]{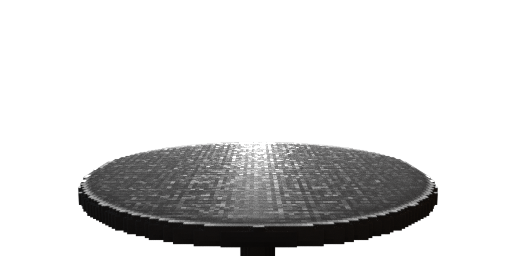}
         \\
        \midrule
        \multirow{2}{*}[+9pt]{\makecell[c]{\rotatebox{90}{\textbf{LoD2}}}}
         & \includegraphics[width=\subFigW,trim={1cm 0cm 1cm 3cm}, clip]{figures/tablev2/mesaSGGX15.png}
         & \includegraphics[width=\subFigW,trim={1cm 0cm 1cm 3cm}, clip]{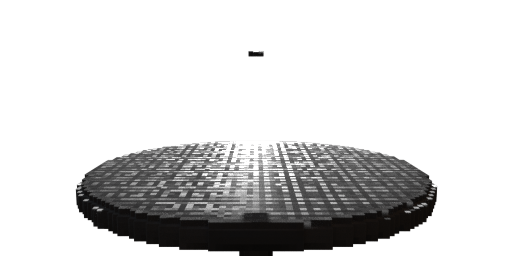}  
         & \includegraphics[width=\subFigW,trim={1cm 0cm 1cm 3cm}, clip]{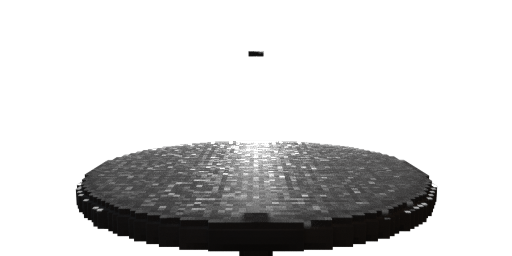}
         \\
        \midrule
        \multirow{2}{*}[+9pt]{\makecell[c]{\rotatebox{90}{\textbf{LoD3}}}}
         & \includegraphics[width=\subFigW,trim={1cm 0cm 1cm 3cm}, clip]{figures/tablev2/mesaSGGX15.png}
         & \includegraphics[width=\subFigW,trim={1cm 0cm 1cm 3cm}, clip]{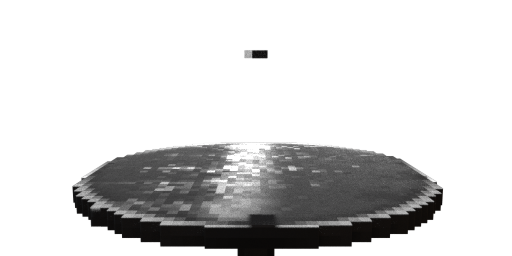}          
         & \includegraphics[width=\subFigW,trim={1cm 0cm 1cm 3cm}, clip]{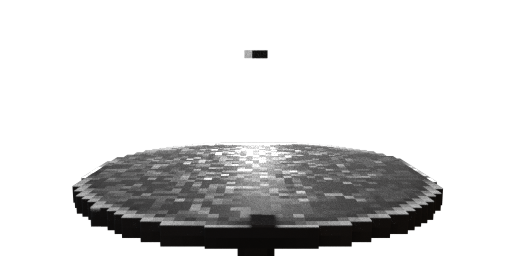}
         \\
        \midrule
    \end{tabular}
    \caption{Render comparison at multiple LoDs of an anisotropic steel table. Left: A three-level resolution voxelized model. Middle: LoD based on naïve average. Right: Our SGGX-H model. In the naïve approach, averages tangent orientations are collapsed into two almost perpendicular directions. Our method evaluates samples of mutliple distributions without averaging them, producing better results, even at coarse resolutions.}
    \label{fig:table-comparisons}
\end{figure}

\label{sec:results}
\subsection{Voxelization pipeline}

We compare our voxelization pipeline to previous rasterized approaches such as Lopez-Moreno et al.~\cite{Lopez-Moreno:2017:Sparse}. We tested our algorithm with a fabric object with more than 20 M of nodes, with 60 spline subdivisions and a resolution of 65536 voxels per side. To test the voxelization pipeline separately, we down-sample the resolution to 8192 voxels by side, selecting the most probable tangent, normal, and a phase function weight with the corresponding identifiers.

We refer to Table~\ref{tab:insertion_times} for voxelization times for three models of a fiber-level fabric patch: up to 160 million nodes were generated and voxelized within a minute in a low-end GPU card in most cases. Sample generations generation are faster in the raster approach. However, we achieve better times for the whole process. The raster approach is suitable for real time applications, while this approach is better for general purposes, big models, and to improve accuracy and sampling control. The raster approach has a hardware limitation due to OpenGL architecture; it cannot render two frames at the same time, while CUDA is more flexible to parallelize the algorithm and to transfer data to and from CPU.

\subsection{Rendering}

To render our generated volumetric data we use a CPU-based volumetric path  with Multiple Importance Sampling (MIS), testing each \lod~using different techniques, and Ground Truth volume rendered at the original volume resolution. For all renders we use Woodcock's Delta Tracking technique~\cite{Woodcock:1965:Techniques} for transmittance estimation.

We show how our method behaves compared to the original data (GT) and na\"ive SGGX fitting for both a fiber-like material~\cite{Khungurn:2016:Matching} using the scarf volume from Jakob et al.~\cite{Jakob:2010:Radiative} (Figure~\ref{fig:scarf-comparisons}); and by voxelizing mesh data along its normal data as orientations (Figures~\ref{fig:table-comparisons},~\ref{fig:helmet-comparisons}).

We also voxelized and rendered vegetation (Figure~\ref{fig:vegetation-comparisons}) and hair (Figure~\ref{fig:hair-comparisons}) models, both known for their difficulty to accurately represent when doing Level of Detail.

To use multiple computed LoDs when rendering, different volumes could be loaded at the same time (one per LoD) and choose which volume to lookup for data at each voxel based in required LoD level. This level can be chosen using usual PathTracer MipMap selection~\cite{Pharr:2016:Physically} based on projected width on camera using the following metric:

\begin{align}
    level   & = MaxLevels - 1 + \log_2(width)                              \\ %
    level_0 & = \lceil level \rceil,\qquad   p_{level_0} = level - level_0 \\
    level_1 & = \lfloor level \rfloor,\qquad p_{level_1} = level_1 - level
\end{align}

\subsection{Quantitative results}

\begin{figure*}[h]
    \centering
    \includegraphics[width=0.9\textwidth, page=2]{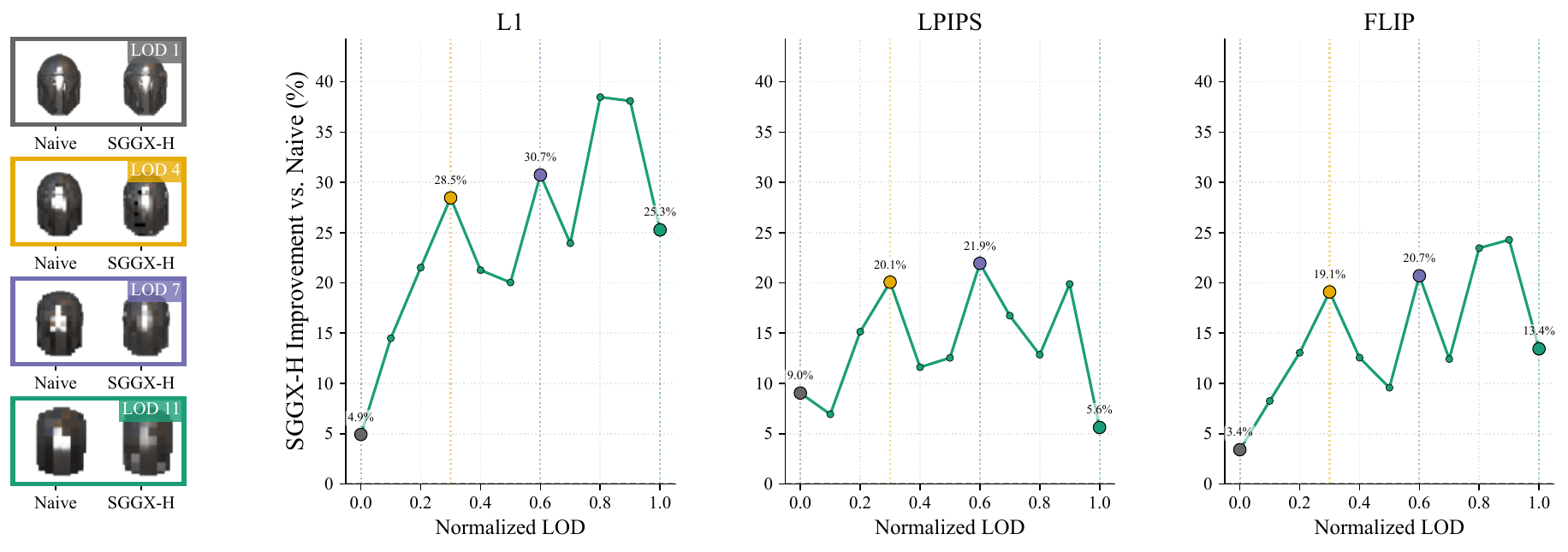}
    \caption{Comparison of SGGX-H relative improvement over Naive rendering for the Helmet model.
             Left column: Side-by-side renderings for selected Levels of Detail (LoDs 1, 4, 7, 11 indicated by colored borders).
             Right columns: Plots showing the percentage improvement of SGGX-H compared to Naive across normalized LoD for L1, LPIPS, and \FLIP error metrics. The x-axis represents normalized LOD, where 0 corresponds to the lowest detail (highest LoD number) and 1 corresponds to the highest detail (lowest LoD number). Positive percentages indicate SGGX-H yields lower error than the Naive approach for the respective metric.}
    \label{fig:helmet_relative_improvement} 
\end{figure*}

In Figure~\ref{fig:helmet_relative_improvement}, we show a quantitative comparison across different LoDs, on relevant metrics, including pixel-wise Mean Absolute Error (L1), as well as perceptual loss (LPIPS~\cite{Zhang:2018:Unreasonable}), and render-aware metric \FLIP~\cite{Andersson:2020:Flip}. We measure the differences on the \textit{Helmet scene} at a constant $256\times256$ resolution, with a mask applied to ignore the background. Across LoD levels, we plot the relative improvement from our hierarchical SGGX aggregation, with respect to the naive aggregation. As shown, our approach consistently outperforms the baseline across all levels of detail. Interestingly, these differences increase as the resolution decreases, particularly on the pixel-wise metrics, suggesting that our approach gains additional benefits as more orientations are aggregated.  \new{We show additional \FLIP metrics for other scenes in Table~\ref{tab:quantitative-metrics}, as well as additional information of each LoD. }

\subsection{Limitations}

Our pipeline exposes three main parameters to the user: \textbf{first level resolution}, \textbf{second level resolution} and \textbf{sample number}. While the former control the final resolution, choosing each one to better adjust to the sparsity of the input model while having enough GPU memory lies on the user. Specifically, combining an elevated number of samples with a high enough resolution can lead also to out of memory problems. While users can control this parameters freely, and desired final resolutions can be achieved by different combinations of resolution parameters, finding an heuristic or metric that suggest suitable combinations of input parameters for a given model would be desirable. 

Additionally, for models where one dimension is almost flat (compared to the largest one), regular cubic voxel shape requires a large increase  of the global resolution so the smaller dimension gets enough nodes to be voxelized properly.
We refer the reader to the Supplementary Material where we discuss a fabric defined by procedural yarns and explicit fiber distributions showing this problematic.

\newcommand{\comparisonRowHelmet}[4]{
  & \ifthenelse{\equal{#1}{}}{\includegraphics[width=\subFigW, height=\subFigH,keepaspectratio]{example-image-duck}}{\includegraphics[width=\subFigW, trim={5cm 3cm 5cm 3cm}, clip]{#1}} 
  & \ifthenelse{\equal{#2}{}}{\includegraphics[width=\subFigW, height=\subFigH,keepaspectratio]{example-image-duck}}{\includegraphics[width=\subFigW, trim={5cm 3cm 5cm 3cm}, clip]{#2}} 
  & \ifthenelse{\equal{#3}{}}{\includegraphics[width=\subFigW, height=\subFigH,keepaspectratio]{example-image-duck}}{\includegraphics[width=\subFigW, trim={5cm 3cm 5cm 3cm}, clip]{#3}} 
  & \ifthenelse{\equal{#4}{}}{\includegraphics[width=\subFigW, height=\subFigH,keepaspectratio]{example-image-duck}}{\includegraphics[width=\subFigW, trim={5cm 3cm 5cm 3cm}, clip]{#4}} 
}

\begin{figure*}[t]
    \newcommand\subFigW{0.16\textwidth}
    \newcommand\subFigH{50px}
    \centering
    \begin{tabular}{
        >{\centering\arraybackslash}m{0.02\textwidth}
        >{\centering\arraybackslash}m{0.02\textwidth}
        >{\centering\arraybackslash}m{0.195\textwidth}
        >{\centering\arraybackslash}m{0.195\textwidth}
        >{\centering\arraybackslash}m{0.195\textwidth}
        >{\centering\arraybackslash}m{0.195\textwidth}}
         &
         & \textbf{Ground Truth}                 
         & \textbf{Averages}                                                                     
         & \textbf{SGGX fit}                                                                     
         & \textbf{Ours} \\
        \midrule
        \multirow{2}{*}[-15pt]{\makecell[c]{\rotatebox{90}{\textbf{Level of detail 1}}}}
         & \rotatebox{90}{\textbf{Renders}}      
         \comparisonRowHelmet{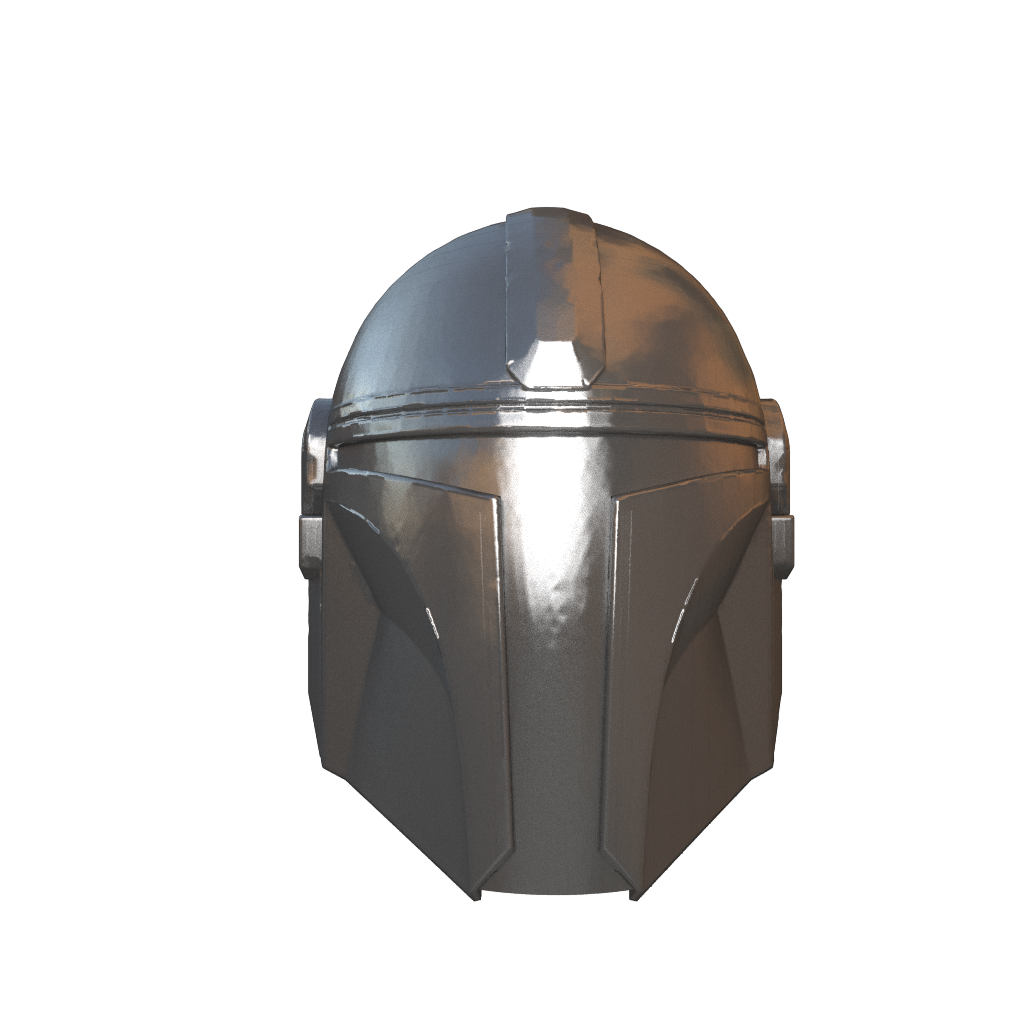}{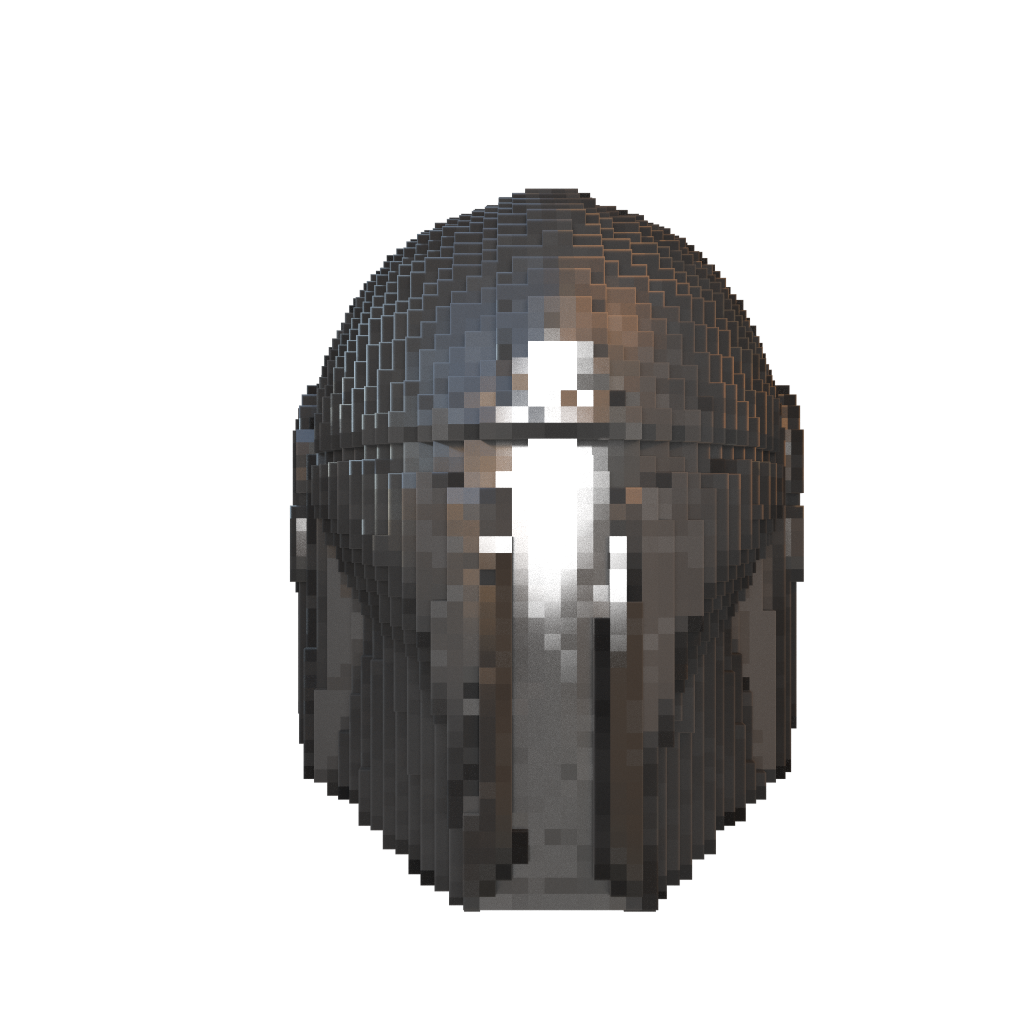}{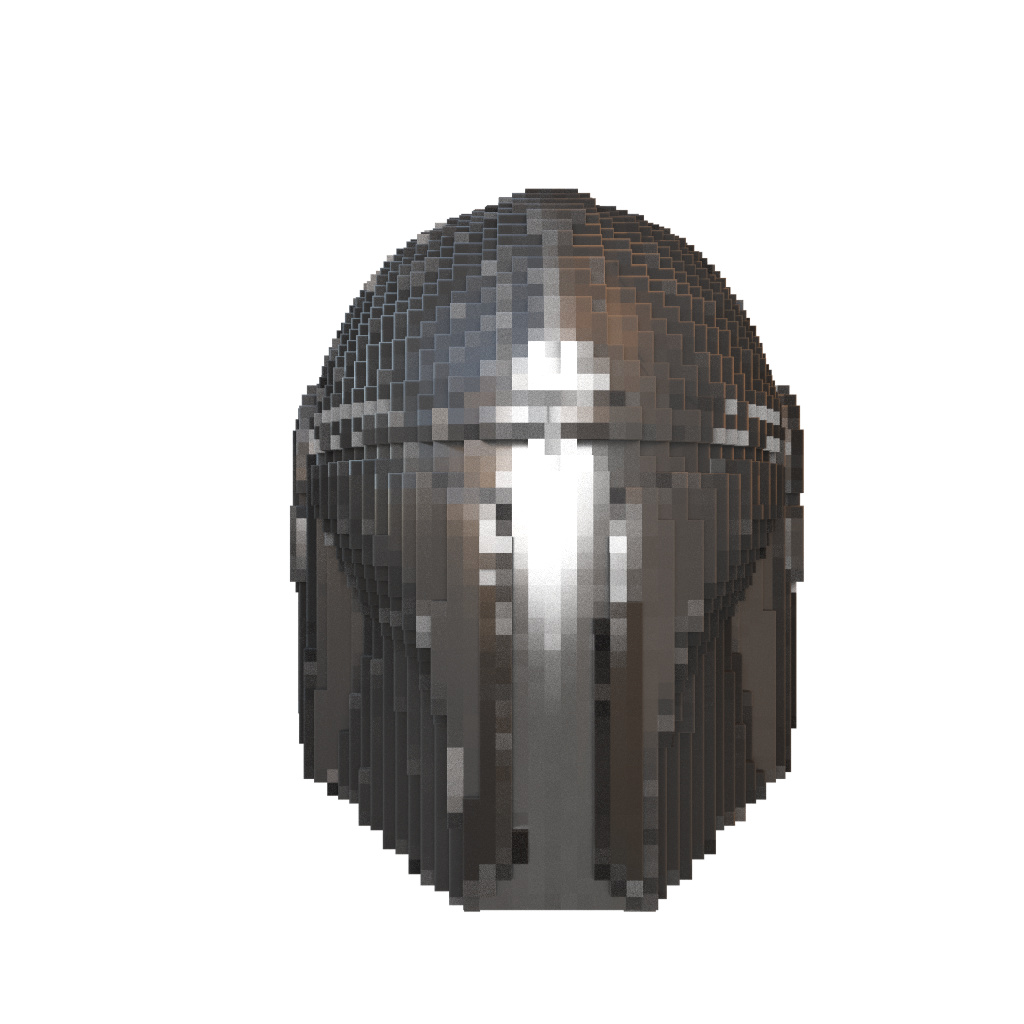}{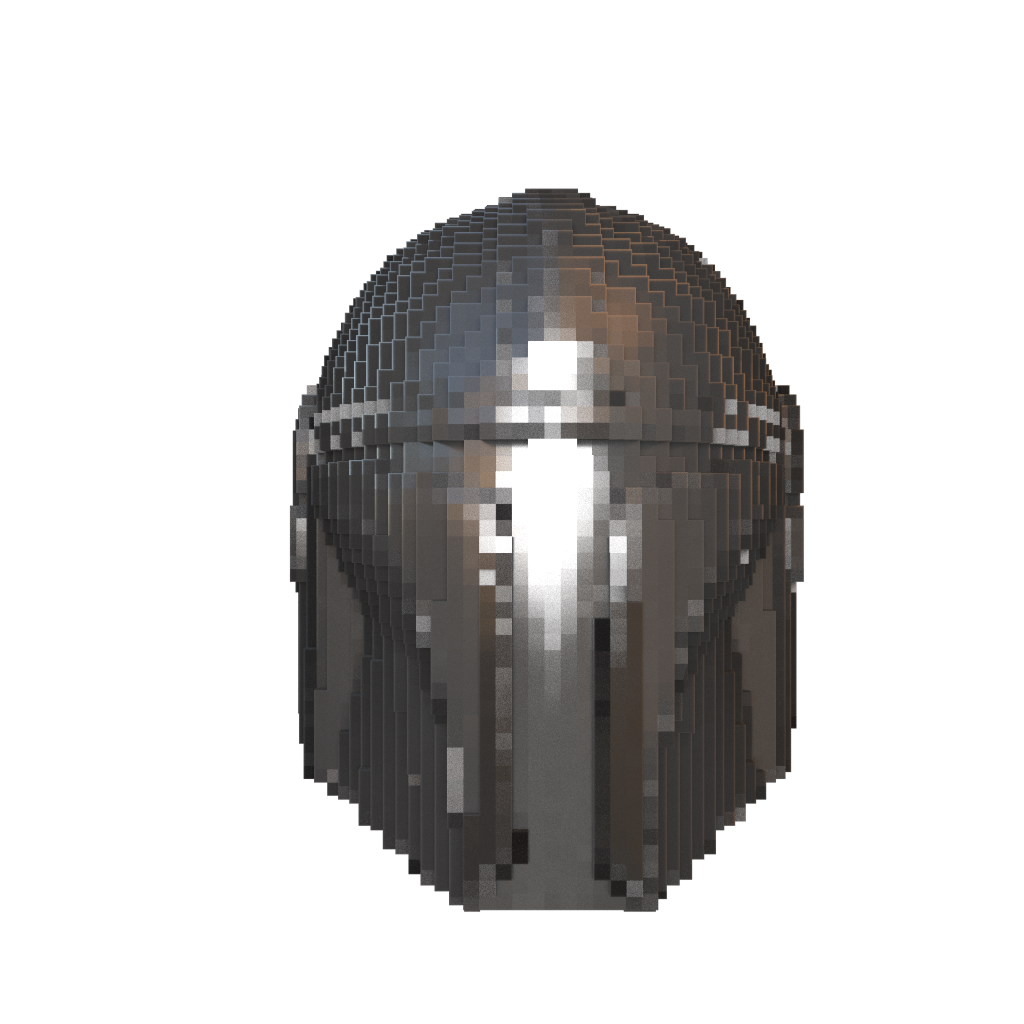}
         \\
         & \rotatebox{90}{\textbf{Orientations}}
         \comparisonRowHelmet{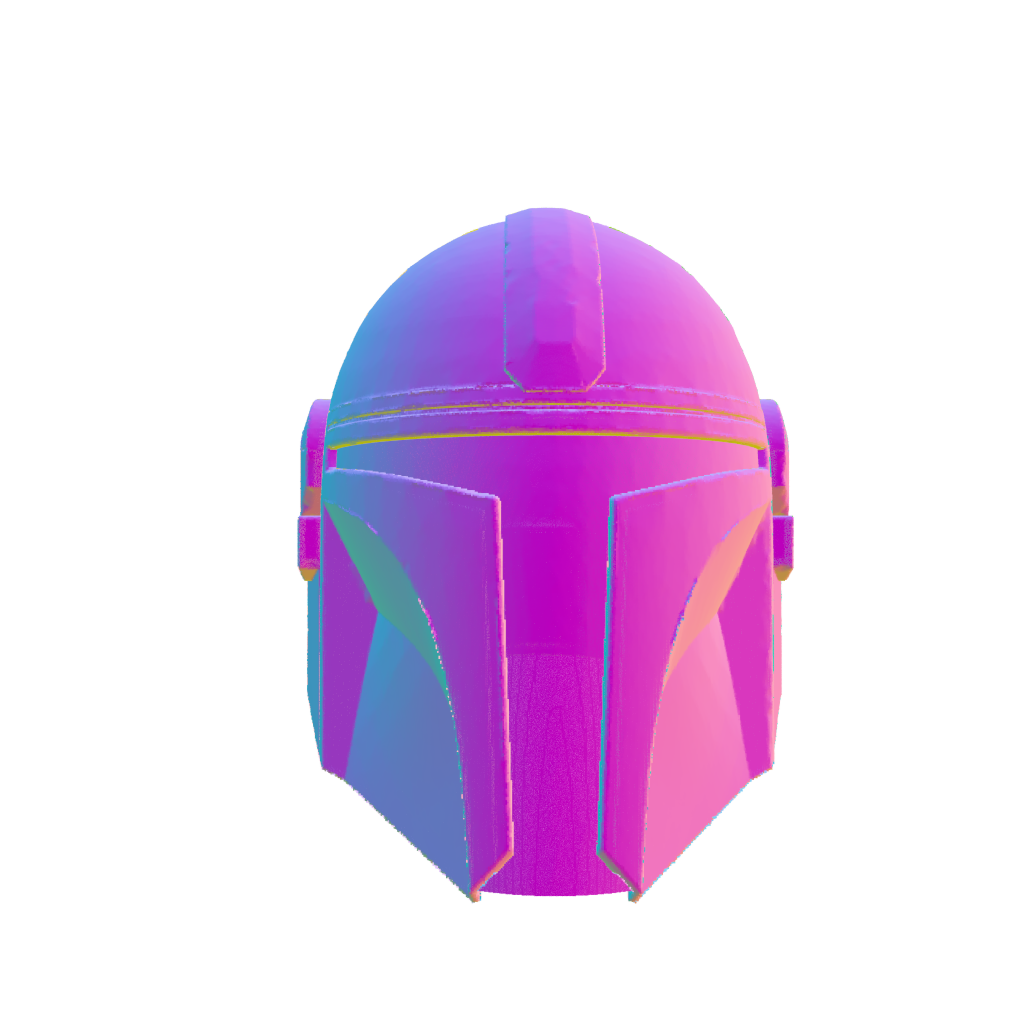}{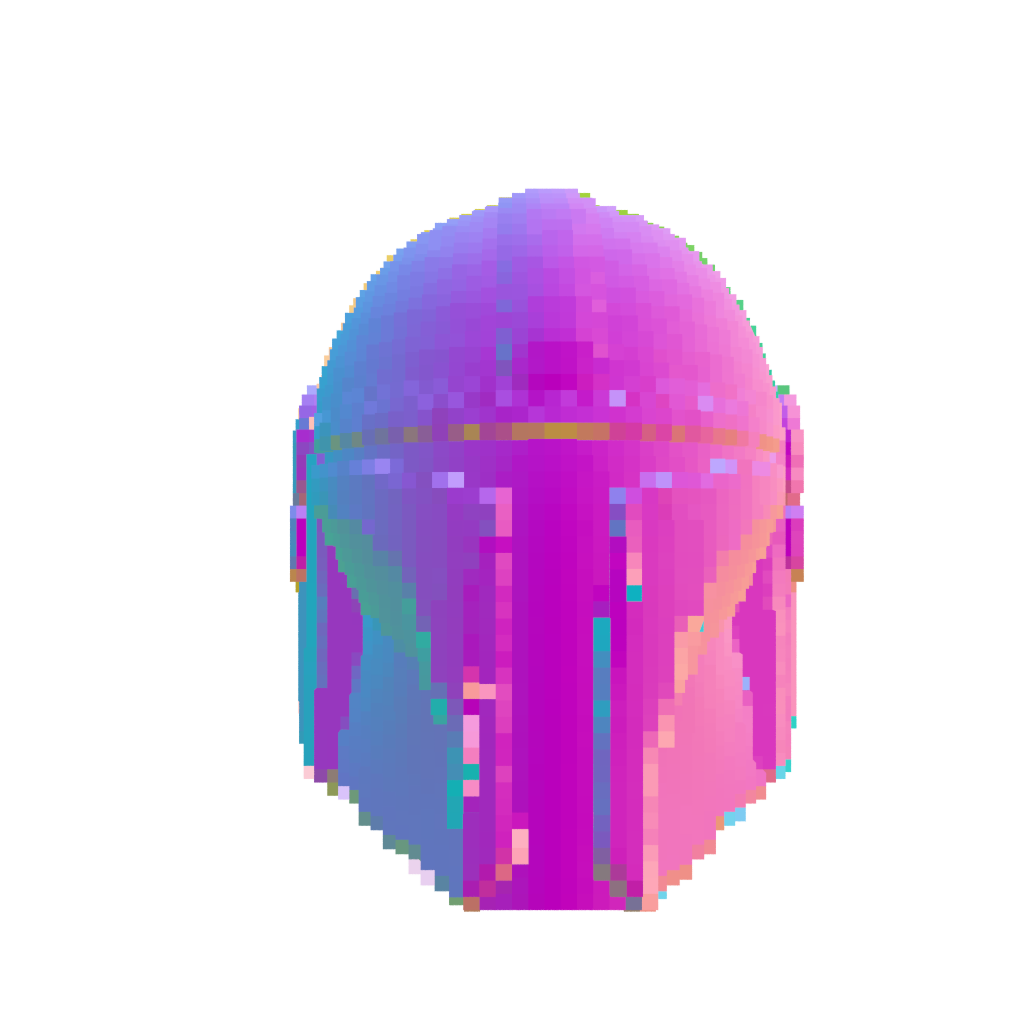}{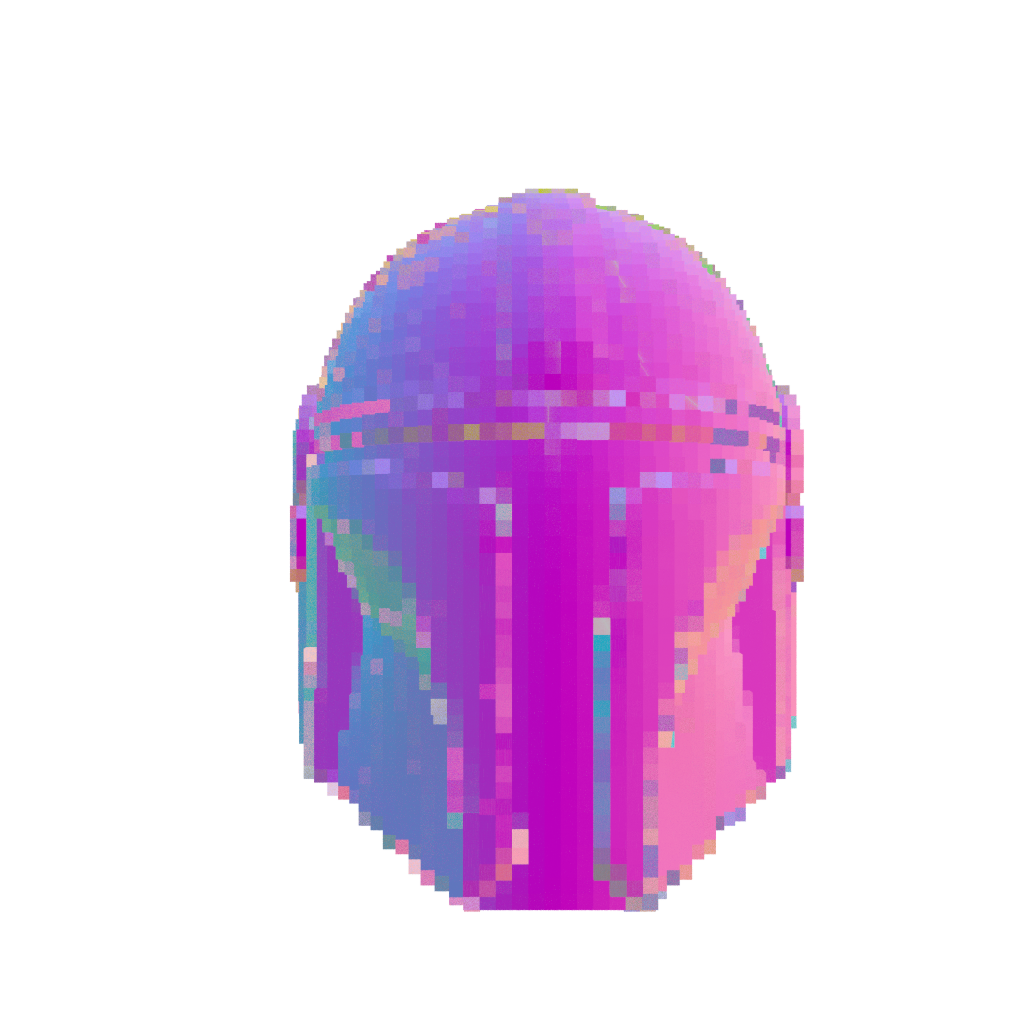}{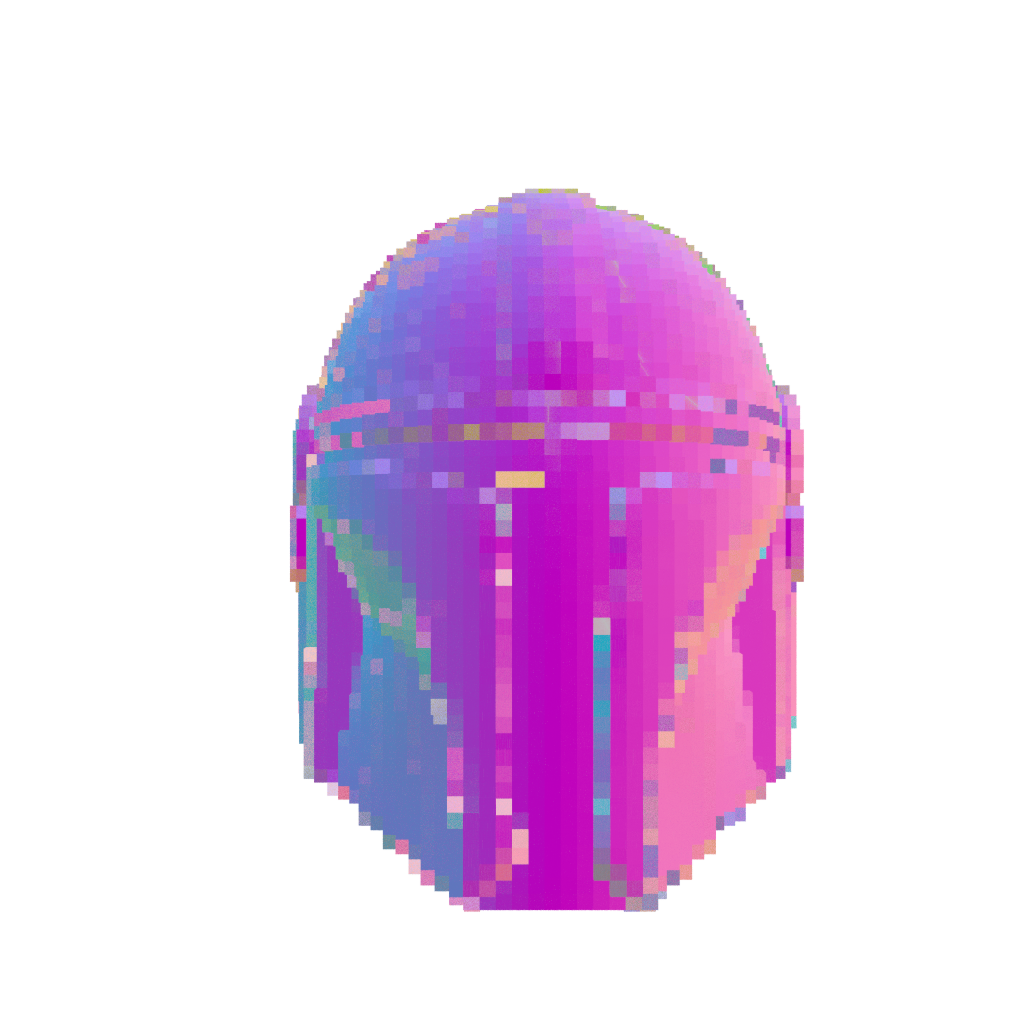}
         \\
        \midrule
        \multirow{2}{*}[-15pt]{\makecell[c]{\rotatebox{90}{\textbf{Level of detail 2}}}}
         & \rotatebox{90}{\textbf{Renders}} 
         \comparisonRowHelmet{figures/helmetv3/cascos-GT.png}{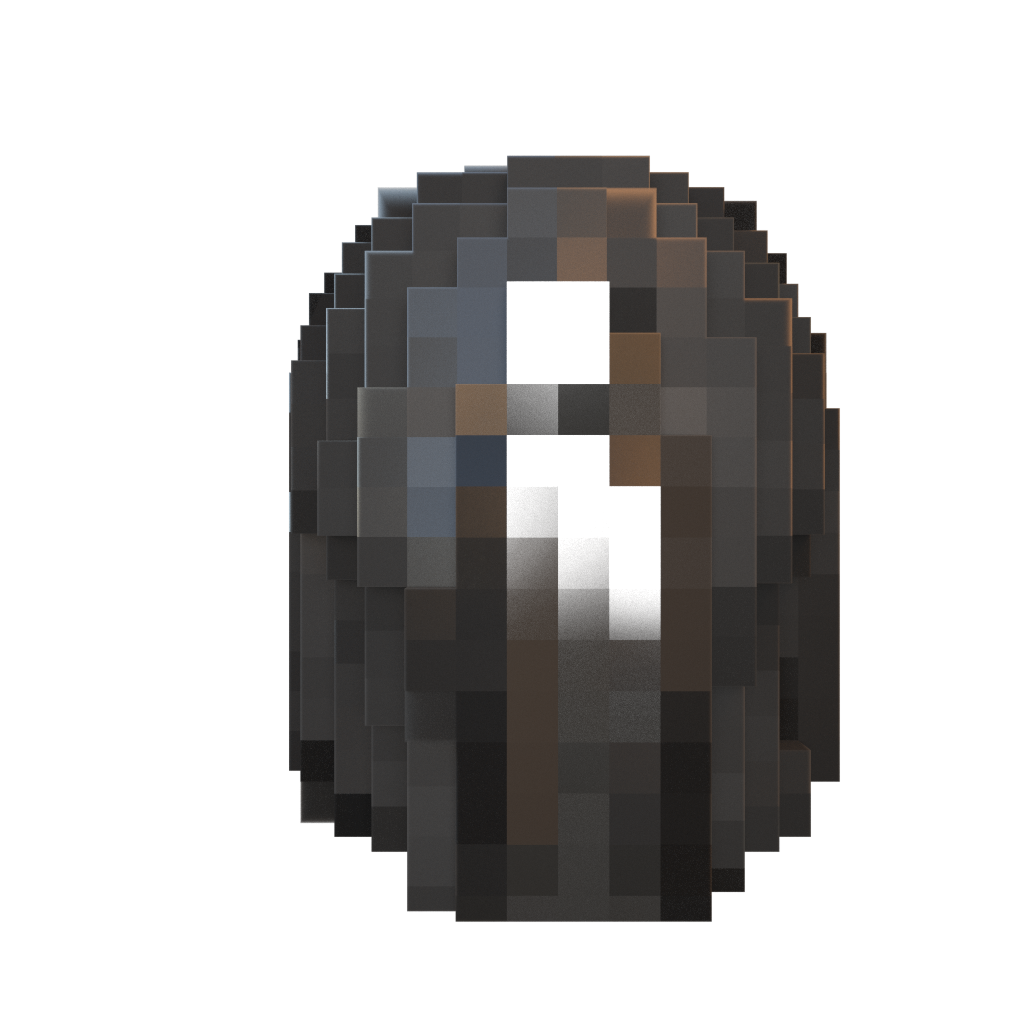}{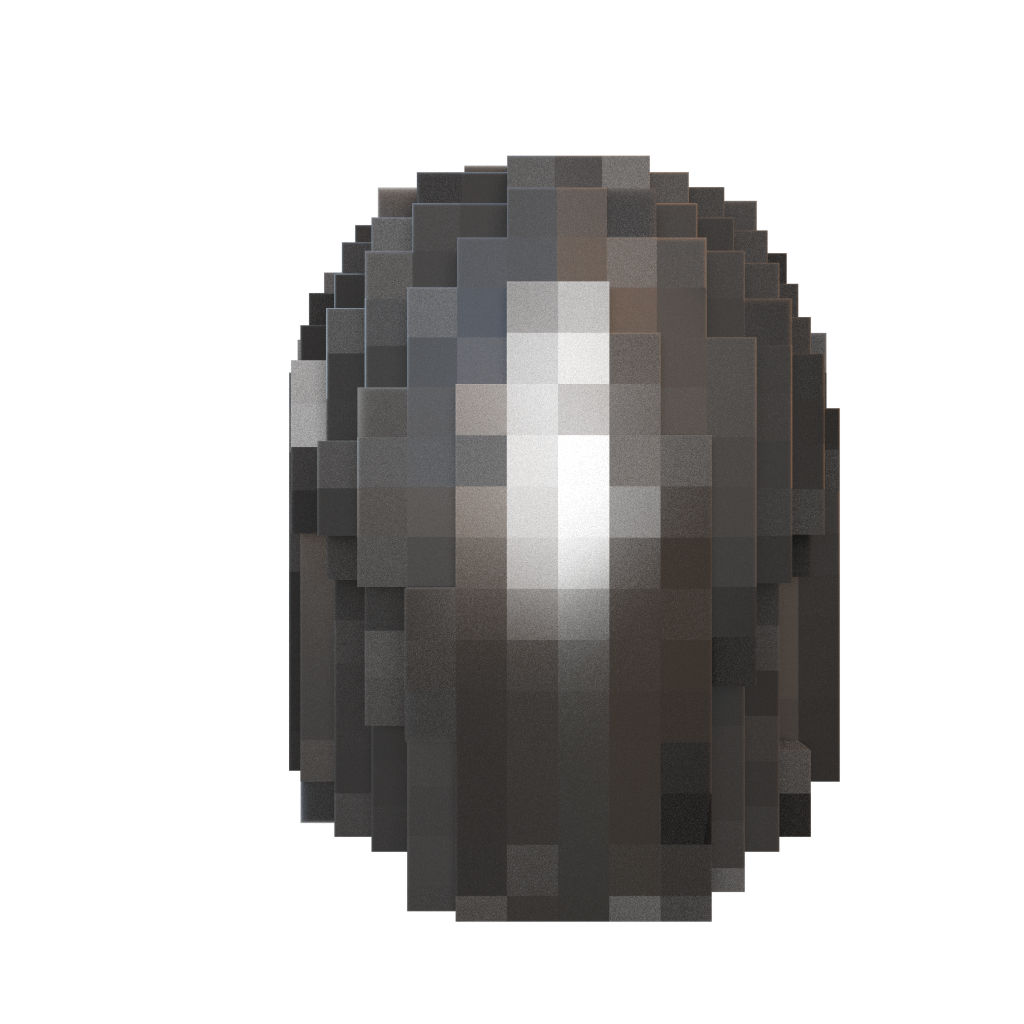}{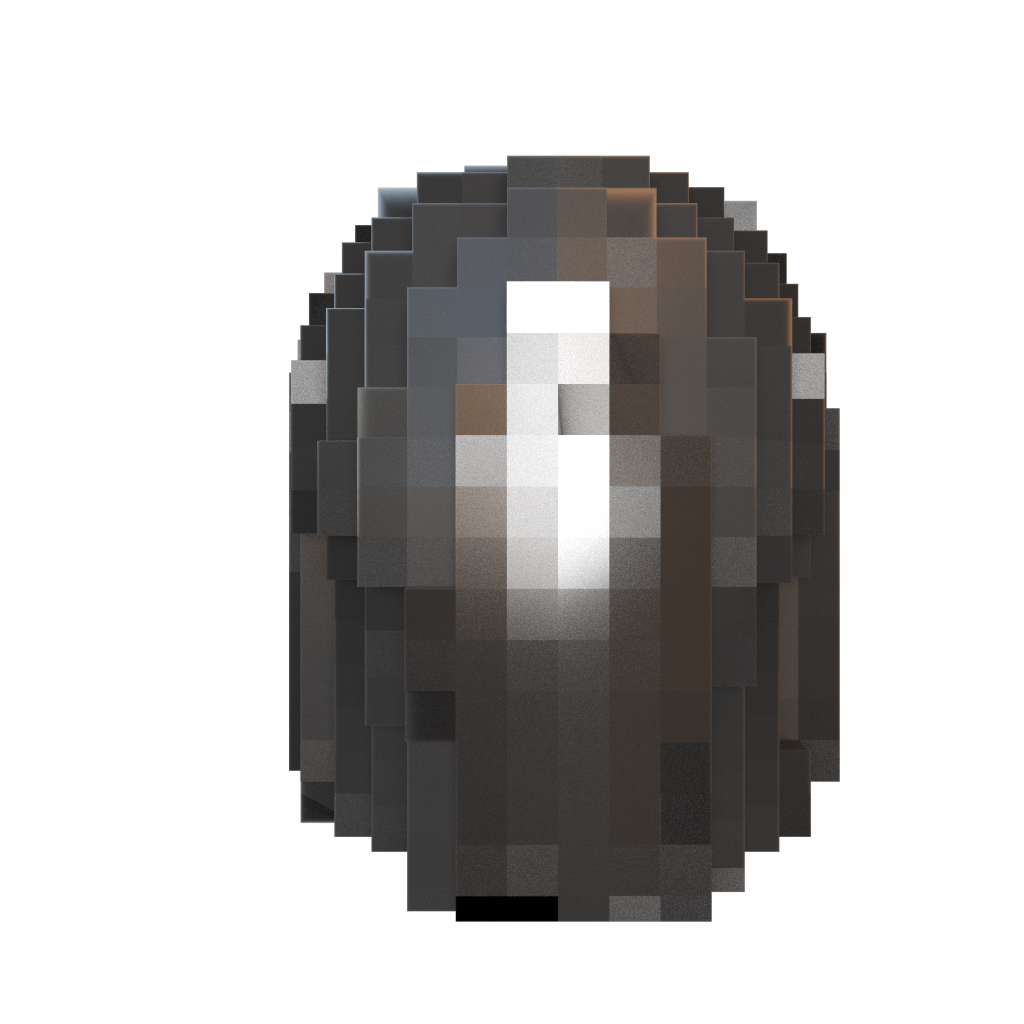} %
         \\
         & \rotatebox{90}{\textbf{Orientations}}
         \comparisonRowHelmet{figures/helmetv3/gt-normals.png}{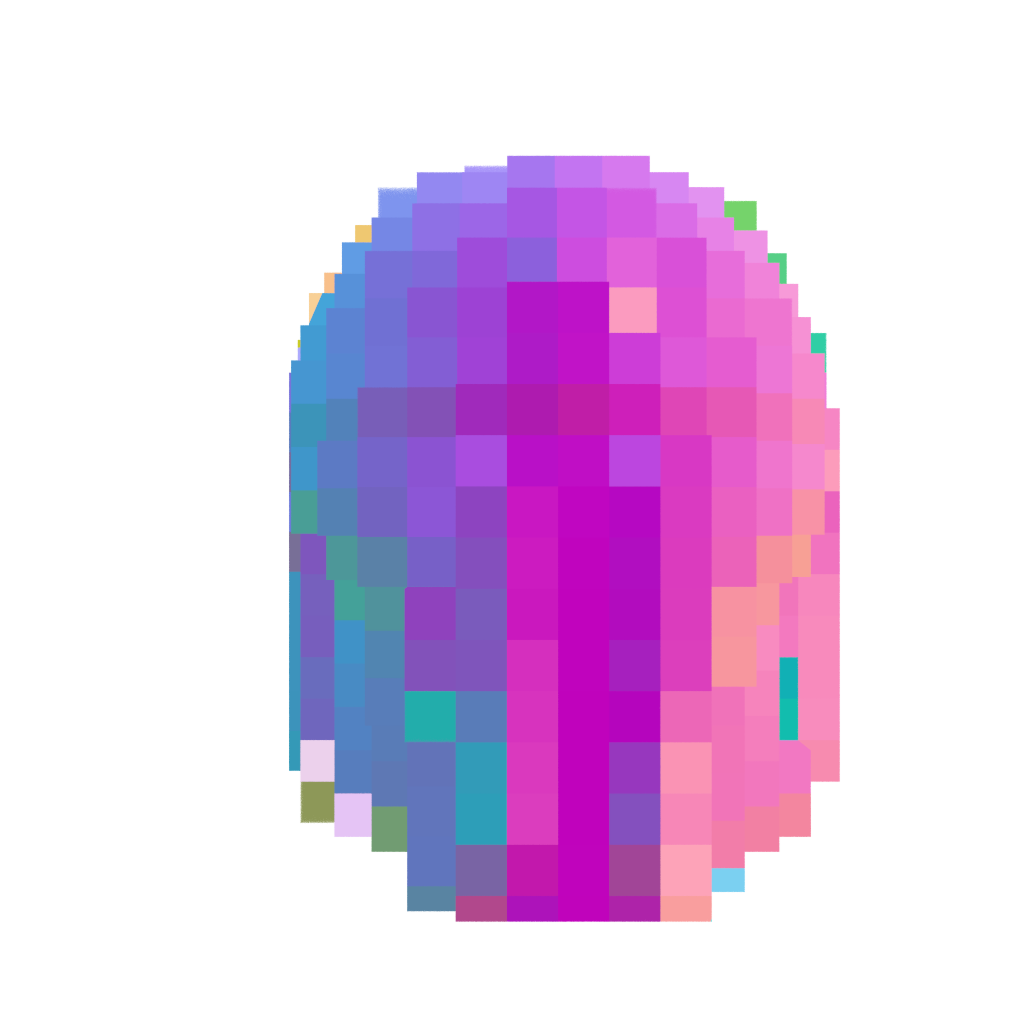}{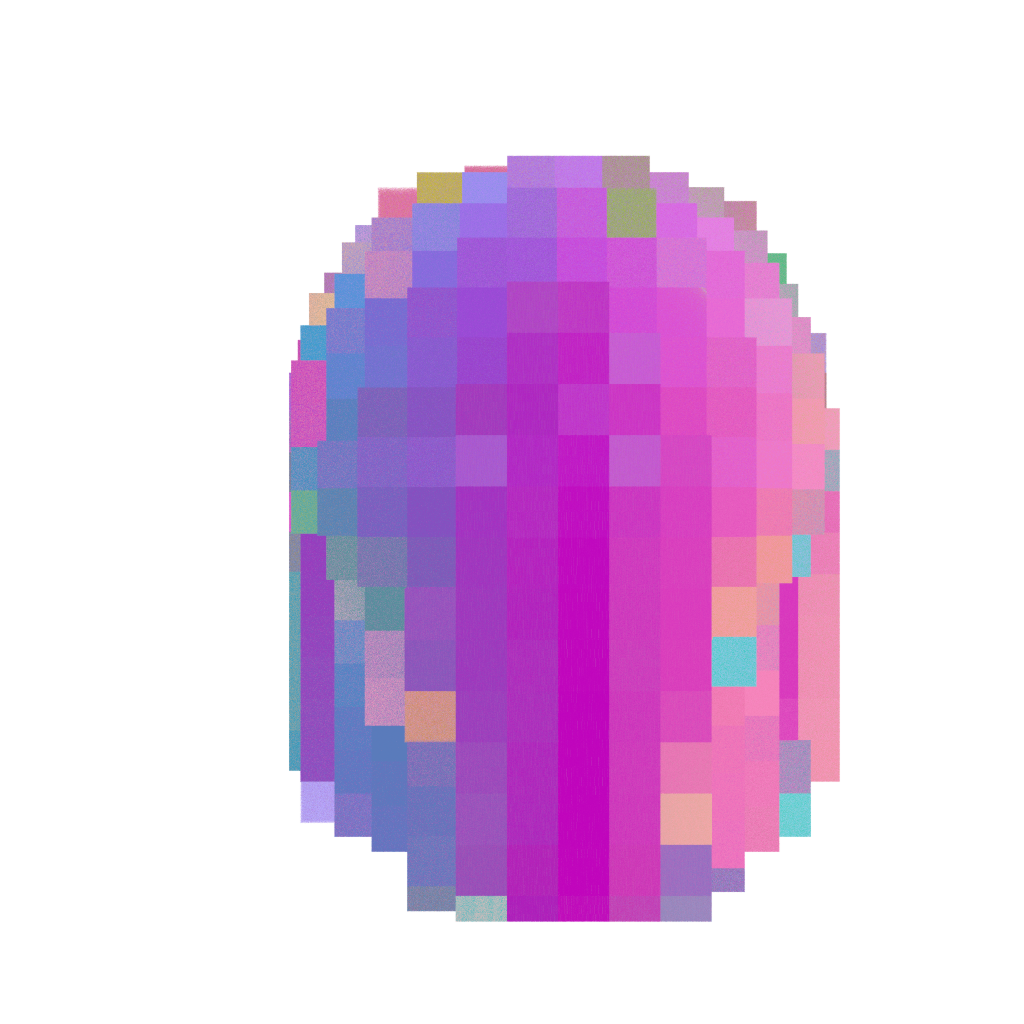}{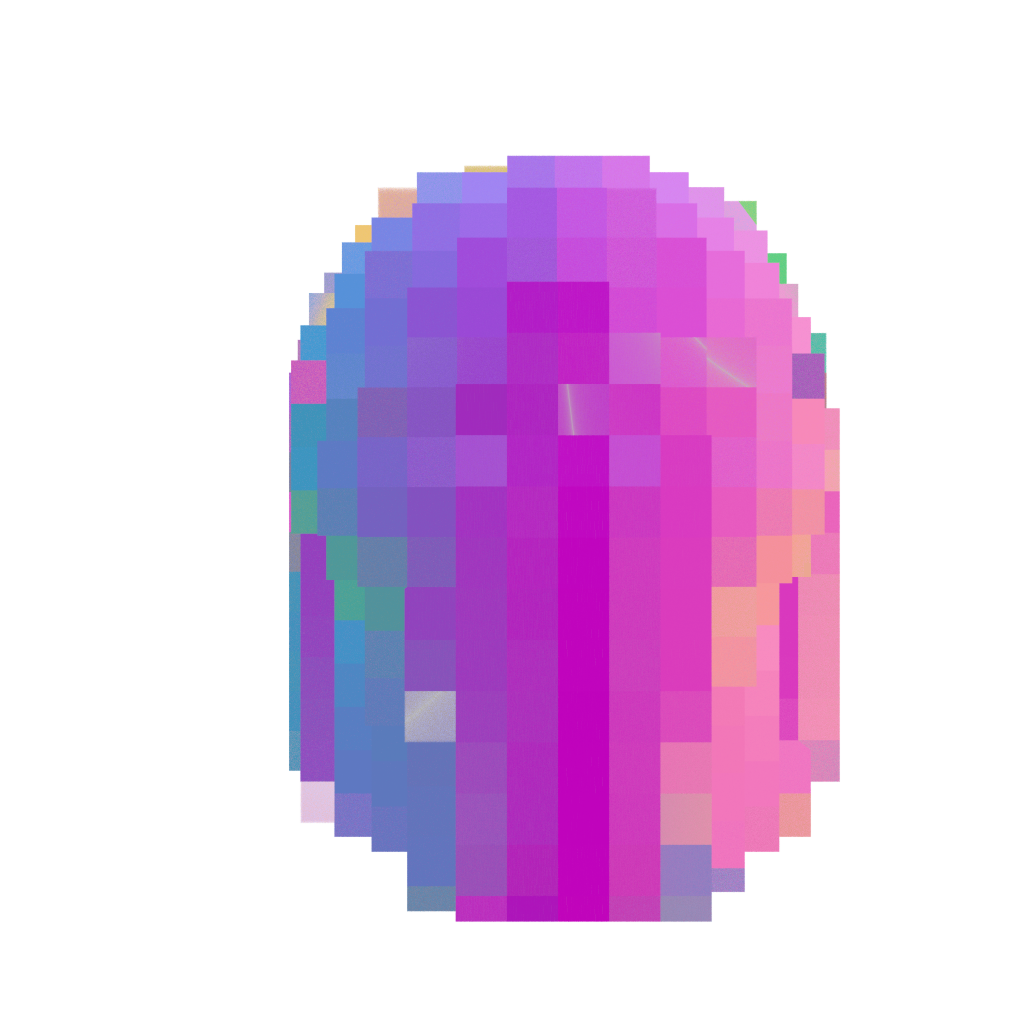}
         \\
        \midrule
        \multirow{2}{*}[-15pt]{\makecell[c]{\rotatebox{90}{\textbf{Level of detail 3}}}}
         & \rotatebox{90}{\textbf{Renders}}
         \comparisonRowHelmet{figures/helmetv3/cascos-GT.png}{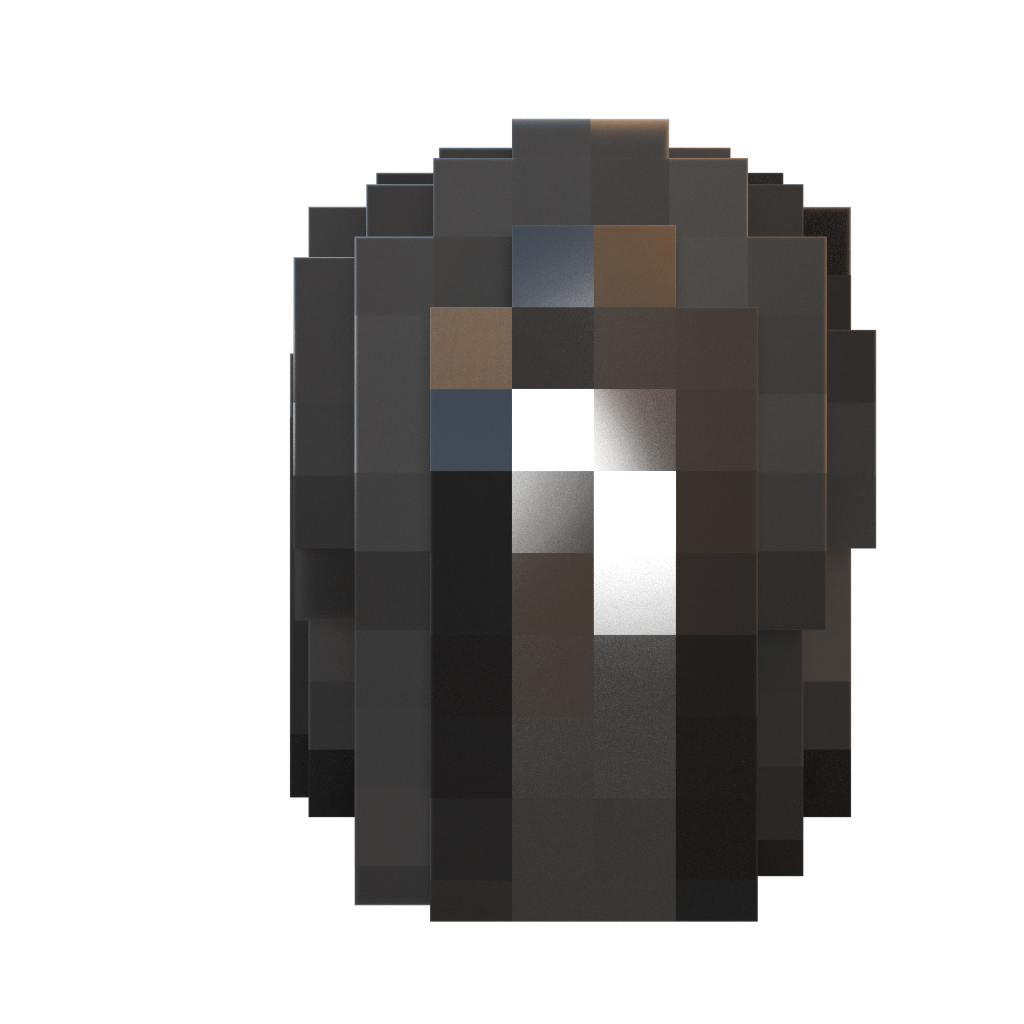}{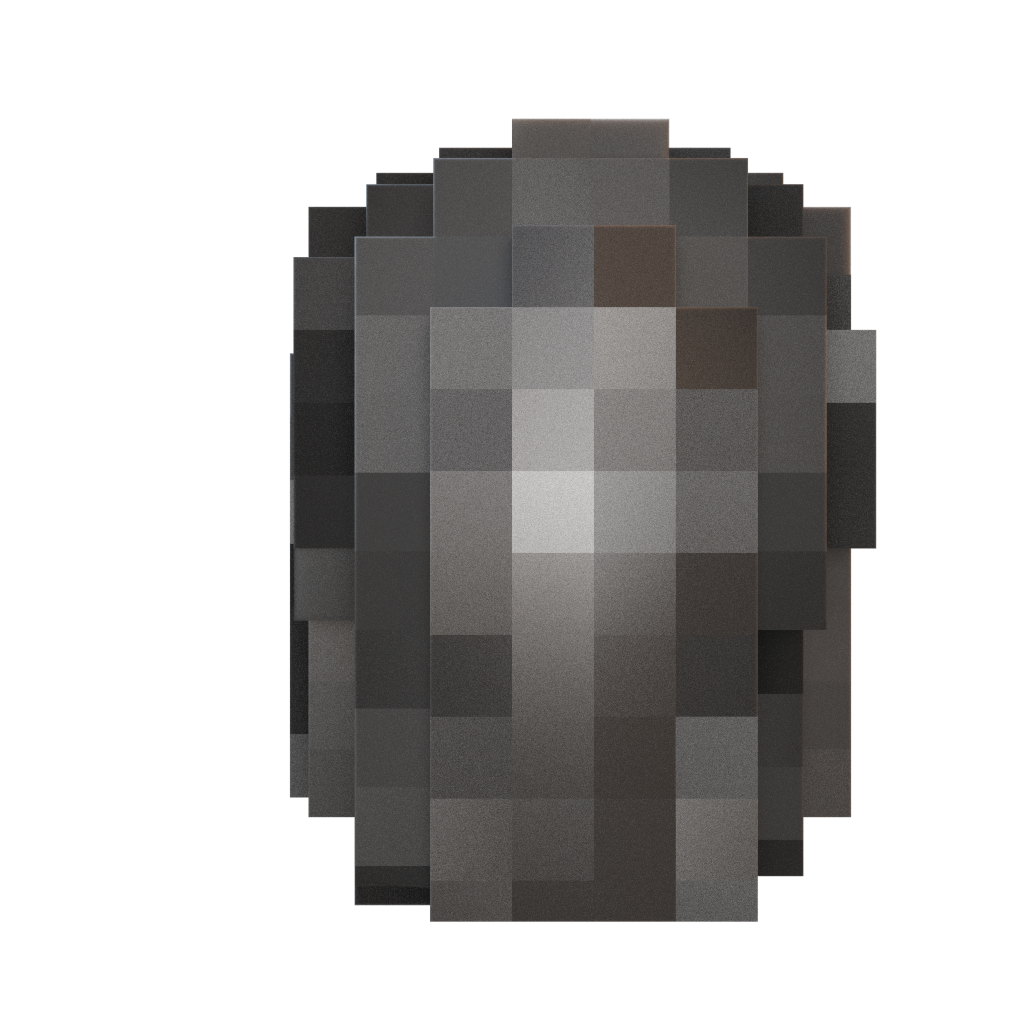}{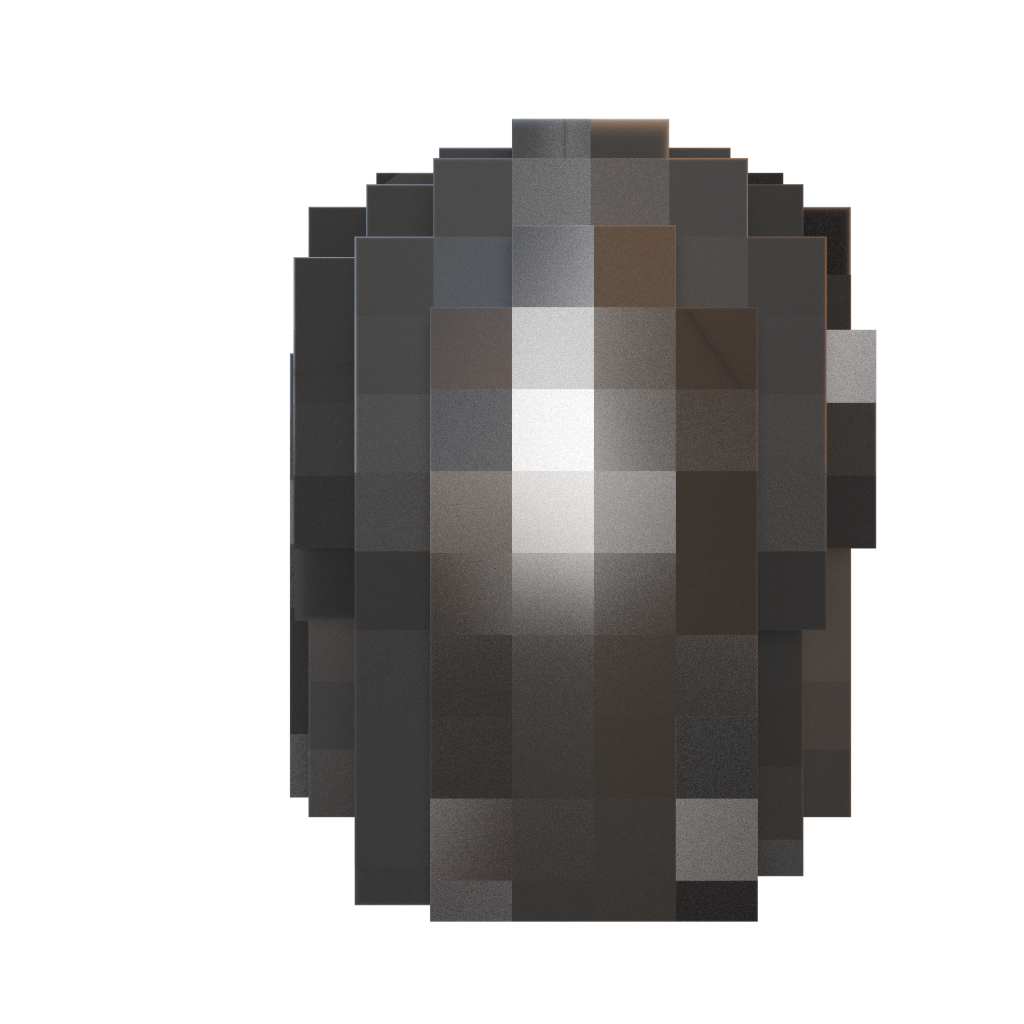} %
         \\
         & \rotatebox{90}{\textbf{Orientations}}
        \comparisonRowHelmet{figures/helmetv3/gt-normals.png}{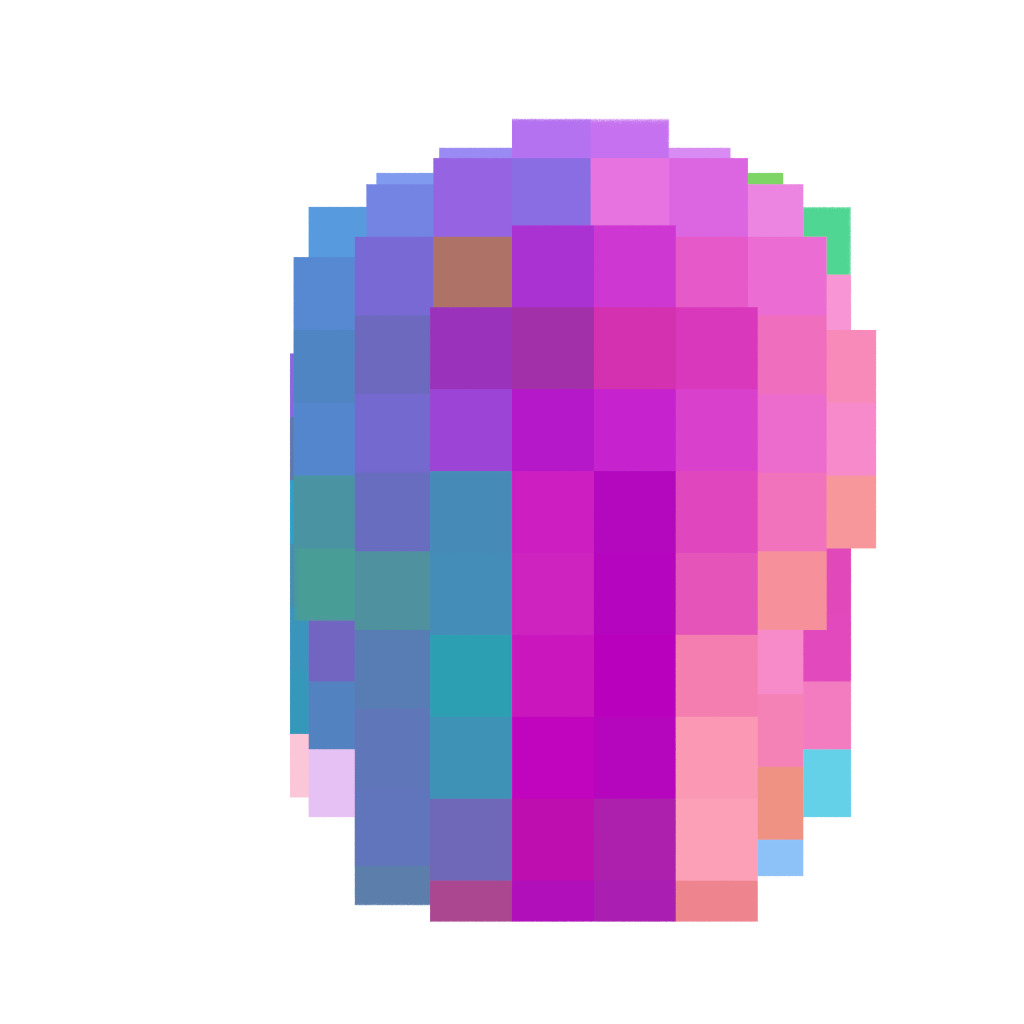}{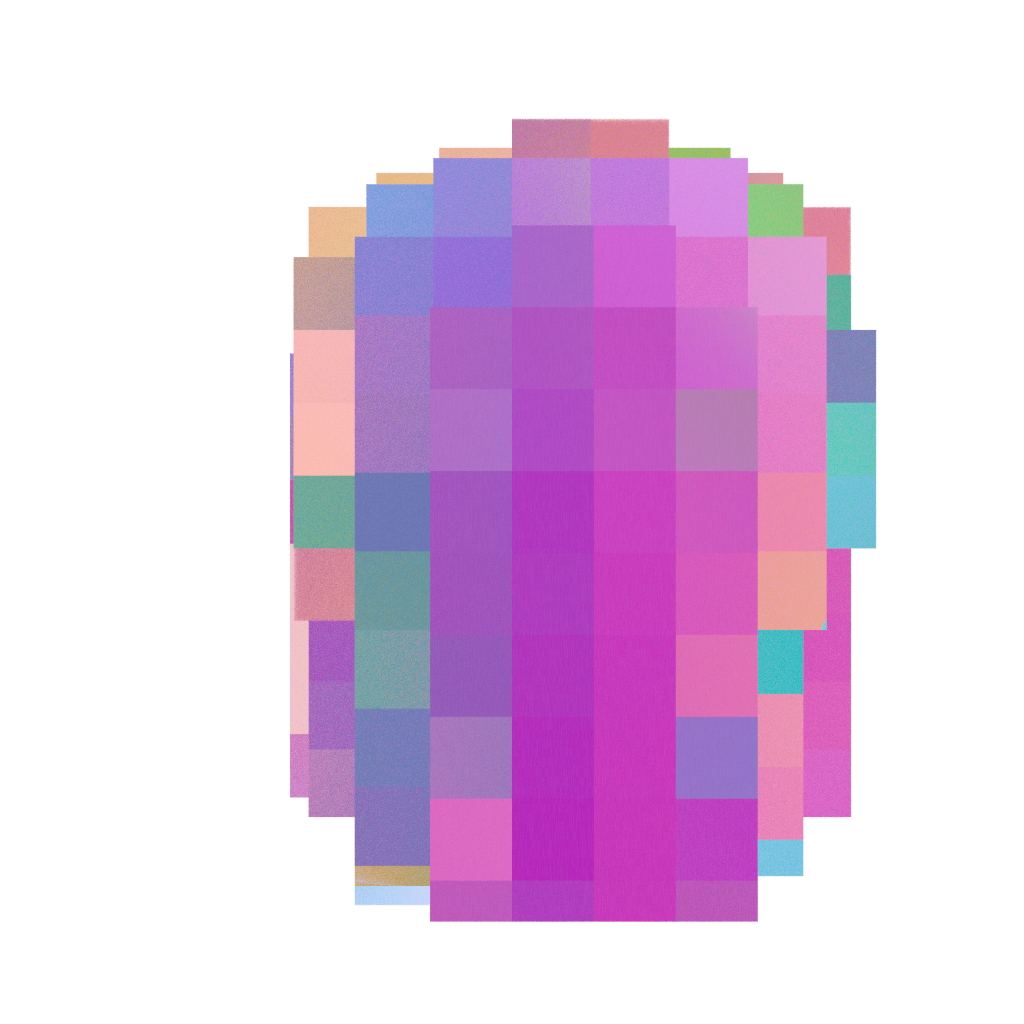}{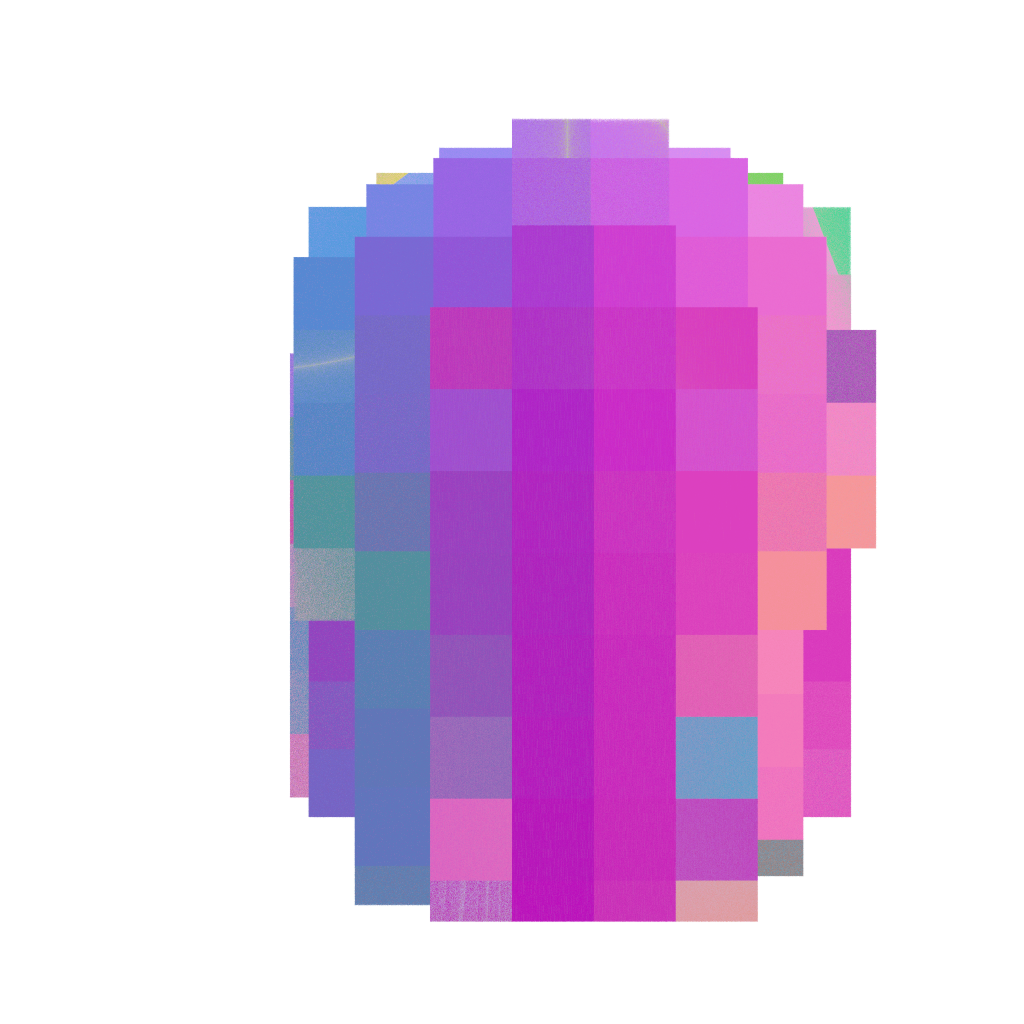}
         \\
    \end{tabular}
    \caption{Comparisons of renders at different voxelization LoDs of a helmet mesh using Ground Truth data (left), naive fit to SGGX model (center) and our hierarchical model (right) for orientation storage. We also show averaged normal per pixel. Note, however, that for both SGGX and SGGX-H we do not evaluate this averaged normal while rendering, but sample normals from the distribution for each shading evaluation.}
    \label{fig:helmet-comparisons}
\end{figure*}

\begin{table}[hb]
    \centering
    \resizebox{.98\columnwidth}{!}{%
        \begin{tabular}{lcccc}
            \toprule 
            \multirow{2}{*}[-1pt]{Scene (Resolution)} & Level of Detail & \multicolumn{3}{c}{Method} \\
            \cmidrule(lr){3-5}
            & (Resolution) & Naive & SGGX & Ours \\
            \midrule
            \midrule
            \cmidrule(lr){1-5}
            \multirow{3}{*}[-1pt]{Table (2048)} %
                & 1 (256) & 0.095   & 0.094  & 0.085 \\
                & 2 (128) & 0.105   & 0.103  & 0.094 \\
                & 3 (64) & 0.111    & 0.118  & 0.101 \\
            \cmidrule(lr){1-5}
            \cmidrule(lr){1-5}
            \multirow{3}{*}[-1pt]{Helmet (512)} %
                & 1 (64) & 0.100  & 0.103  & 0.098 \\
                & 2 (32) & 0.163  & 0.154  & 0.151 \\
                & 3 (16) & 0.201  & 0.181  & 0.179 \\
            \cmidrule(lr){1-5}
            \multirow{3}{*}[-1pt]{Scarf ($\approx$1024)} %
                & 1 ($\approx$800) & 0.0642      & 0.091   & 0.0641 \\
                & 2 ($\approx$400) & 0.103      & 0.133   & 0.101 \\
                & 3 ($\approx$200) & 0.281      & 0.284   & 0.202 \\
            \bottomrule
            \bottomrule
        \end{tabular}%
    }
    \caption[Quantitative metrics]{While a straightforward SGGX fit sometimes fail to fully capture aggregated our method (SGG-H) scores improved (\FLIP) metrics compared to a Naive approach for all scenes, at different Levels of Detail. }
    \label{tab:quantitative-metrics}
\end{table}

\section{Conclusions and Future Work}
\label{sec:future-work}

In this paper, we have introduced a novel voxelization method, optimized to deal with sparse microgeometry while maintaining memory efficiency. In addition to being able to efficiently handle sparse data, our method allows voxelization of complex data along density. Furthermore, it can fit distributions to generate lower resolution volumes, suitable for LoD, that accurately represent not only \textit{ocupancy} and \textit{optical density} but also orientation data such as normals or tangents. Our work can be extended in several directions.

In section~\ref{sec:processing:density}, we described the LoD approach for density data. Due to its limited per-voxel accuracy with complex geometries, this method could be improved by exploring novel neural or statistical representations. These representations could encode aggregated density, allowing for sampling and combination in a way similar to our solution for orientation data. Additionally, such approach should help removing artifacts like the ones present when voxelizing complex fabrics.

Regarding distribution merging, method uses a hierarchical approach and aims to reduce the amount of SGGXs stored per node, however, we still need to re--fit some of them, possibly reducing the accuracy of the orientation distribution for that node. 

Ideally, we could use all parent nodes to generate a mix of distributions on--the--fly that would accurately represent the full orientation distribution for a selected child node. However, this comes with the price of exploring, storing and merging all required parent nodes recursively. Future research on how to improve such approach could lead to a higher accuracy on sparse LoDs (hence rendering quality), while saving over all space when storing all LoDs together, avoiding orientation data duplication.

Also, the possible benefits of alternative GPU-CPU data flows could be analyzed. For instance, in our implementation we fit the leaf node SGGX in GPU, but then we generate an histogram using the SGGX itself in CPU to build the SGGX-H representation. While this is necessary for aggregated SGGX distributions, for leaf nodes we could use the original GPU-computed histogram instead of sampling, at the cost of moving that data from GPU to CPU. Newer GPU features could be explored in the future to try to overcome these issues.

\begin{table*}[hb]
    \centering
    \resizebox{\textwidth}{!}{
        \begin{tabular}{cccccccc}
            \toprule
            \multirow{2}{*}{Scene}         & \multicolumn{4}{c}{Parameters} & \multicolumn{3}{c}{Time (s)}                                                                                             \\ \cmidrule(lr){2-5} \cmidrule(lr){6-8}
                                           & Nodes                          & Fibers                       & Subdivisions        & Resolution (1st x 2nd) & Raster                         & Ours   & w/ Histogram \\ \hline \hline
            \multirow{3}{*}{Fabric flat}         & \multirow{3}{*}{8,359,496}     & \multirow{9}{*}{150}         & \multirow{9}{*}{60} & 32 x 32    & 81.117                         & 3.450  & 8.813        \\
                                           &                                &                              &                     & 32 x 64    & 15.677                         & 6.058  & 9.151        \\
                                           &                                &                              &                     & 32 x 128   & 41.996                         & 22.339 & 9.396        \\ \cmidrule(lr){1-2} \cmidrule(lr){5-8}
            \multirow{3}{*}{Fabric folded} & \multirow{3}{*}{8,227,286}     &                              &                     & 32 x 32    & 6.649                          & 3.521  & 7.987        \\
                                           &                                &                              &                     & 32 x 64    & 14.502                         & 6.156  & 8.2062       \\
                                           &                                &                              &                     & 32 x 128   & 30.796                         & 21.152 & 58.188       \\ \cmidrule(lr){1-2} \cmidrule(lr){5-8}
            \multirow{3}{*}{Fabric Hanging}       & \multirow{3}{*}{16,3237,089}   &                              &                     & 32 x 32    & \multirow{3}{*}{Out of memory} & 20.485 & 58.188       \\
                                           &                                &                              &                     & 32 x 64    &                                & 22.346 & 66.106       \\
                                           &                                &                              &                     & 32 x 128   &                                & 36.113 & 48.643       \\
            \bottomrule
        \end{tabular}
    }
    \caption[Voxelization by insertion]{Using our voxelizer in CUDA, we measured the times to replicate the behavior of previous approaches. Measured time include loading stage, voxelization, and exportation to OpenVDB format. We compare our voxelization times with and without histogram computation against a raster based implementation using the same GPU (NVIDIA GeForce - GTX 1080 Ti). Visualization of the fabrics voxelized for these test are shown in the Supplementary Material. }
    \label{tab:insertion_times}
\end{table*}

\section*{Declarations}
\subsection*{Funding}
This publication is part of the project TaiLOR, CPP2021-008842 funded by MCIN/AEI/10.13039/501100011033 and the NextGenerationEU / PRTR programs.

\bibliographystyle{ieeetr}
\bibliography{strings-full, rendering-bibtex, additional}   %

\end{document}